\documentclass[a4paper, 10pt]{article}
\usepackage{fullpage}

\newcommand\bigone[1]{}
\newcommand\smallone[1]{#1}
\usepackage{amsfonts,amssymb,amsmath,amsthm,latexsym}
\usepackage{array,xspace}

\newcommand{\ignore}[1]{}

\newcommand{\ee}{\mathsf{e}}
\newcommand{\eps}{\varepsilon}

\newcommand{\tr}{\mathop{\mathrm{tr}}}

\newcommand{\etal}{{\em et al.}\xspace}

\def\makeletter#1{%
\expandafter \newcommand \csname b#1\endcsname {\mathbb{#1}}%
\expandafter \newcommand \csname c#1\endcsname {\mathcal{#1}}%
\expandafter \newcommand \csname t#1\endcsname {\widetilde{#1}}%
\expandafter \newcommand \csname ct#1\endcsname {\widetilde{\mathcal{#1}}}%
}
\def\makeletters(#1#2){\makeletter#1\ifx#2.\else\makeletters(#2)\fi}
\makeletters(QWERTYUIOPASDFGHJKLZXCVBNM.)

\def\makeSkob#1#2#3{%
\def\LLL{\left} \def\RRR{\right}
\expandafter \edef \csname #1\endcsname #2##1#3{\SkobInner}
\def\LLL{\bigl} \def\RRR{\bigr}
\expandafter \edef \csname #1A\endcsname #2##1#3{\SkobInner}
\def\LLL{\Bigl} \def\RRR{\Bigr}
\expandafter \edef \csname #1B\endcsname #2##1#3{\SkobInner}
\def\LLL{\biggl} \def\RRR{\biggr}
\expandafter \edef \csname #1C\endcsname #2##1#3{\SkobInner}
\def\LLL{\Biggl} \def\RRR{\Biggr}
\expandafter \edef \csname #1D\endcsname #2##1#3{\SkobInner}
\def\LLL{} \def\RRR{}
\expandafter \edef \csname #1O\endcsname #2##1#3{\SkobInner}
}

\def\SkobInner{\LLL(##1\RRR)} \makeSkob{s}[]
\def\SkobInner{\LLL[##1\RRR]} \makeSkob{sk}[]
\def\SkobInner{\LLL\lbrace##1\RRR\rbrace} \makeSkob{sfig}{}{}
\def\SkobInner{\LLL\lfloor##1\RRR\rfloor} \makeSkob{floor}[]
\def\SkobInner{\LLL\lceil##1\RRR\rceil} \makeSkob{ceil}[]
\def\SkobInner{\LLL\langle##1\RRR\rangle} \makeSkob{ip}<>
\def\SkobInner{\LLL|##1\RRR\rangle} \makeSkob{ket}|>
\def\SkobInner{\LLL|##1\RRR|} \makeSkob{abs}||
\def\SkobInner{\LLL\|##1\RRR\|} \makeSkob{norm}||
\def\SkobInner{\LLL\|##1\RRR\|_{\noexpand\mathrm F}} \makeSkob{normFrob}||
\def\SkobInner{\LLL\|##1\RRR\|_{\noexpand\mathrm{tr}}} \makeSkob{normtr}||

\newcommand{\midA}{\mathbin{\bigl|}}

\def \elem[#1]{[\![#1]\!]}
\def \bigfrac#1/{\left.#1\right/}
\def \bigfracR/#1.{\left/#1\right.}

\newcommand{\pfstart}{\begin{proof}} 
\newcommand{\pfsketch}{\begin{proof}[Proof sketch]}
\newcommand{\pfend}{\end{proof}} 
\newcommand{\itemstart}{\begin{itemize}\itemsep0pt}
\newcommand{\itemend}{\end{itemize}}
\newcommand{\descrstart}{\begin{description}\itemsep0pt}
\newcommand{\descrend}{\end{description}}
\newcommand{\enumstart}{\begin{enumerate}\itemsep0pt}
\newcommand{\enumend}{\end{enumerate}}

\newcommand{\CrossRef}[2]{#1~\ref{#2}}

\bigone{\newtheorem{thm}{Theorem}[chapter]}
\smallone{\newtheorem{thm}{Theorem}}

\newcommand{\maketheorem}[2]{
\newtheorem{#1}[thm]{#2}
\expandafter\def \csname ref#1\endcsname ##1{\CrossRef{#2}{#1:##1}}
}

\maketheorem{lem}{Lemma}
\maketheorem{prp}{Proposition}
\maketheorem{cor}{Corollary}
\maketheorem{clm}{Claim}
\maketheorem{fact}{Fact}

\theoremstyle{definition}
\maketheorem{rem}{Remark}
\maketheorem{obs}{Observation}
\maketheorem{defn}{Definition}
\maketheorem{exm}{Example}
\maketheorem{assump}{Assumption}

\def \rf(#1:#2){\csname ref#1\endcsname{#2}}

\def\mycommand#1#2{
\expandafter\newcommand \csname#1\endcsname {#2}%
}
\def\remycommand#1#2{
\expandafter\renewcommand \csname#1\endcsname {#2}%
}

\usepackage{algorithm}

\newcommand\draft[1]{}
\newcommand\release[1]{#1}
\release{\usepackage[breaklinks]{hyperref}}

\newcommand{\Adv}{\mathop{\mathrm{ADV}^{\pm}}}
\newcommand{\pAdv}{\mathop{\mathrm{ADV}}}
\newcommand{\pr}{\mathop{\mathrm{Pr}}}
\newcommand{\dd}{\mathrm{d}}

\newcommand{\reg}[1]{{\mathsf{#1}}}

\def \ket#1|#2>%
{\ifx&#1&
|#2\rangle
\else
|#2\rangle_{\mathsf #1}
\fi}

\title{Quantum Algorithms for Learning Symmetric Juntas via the Adversary Bound}
\author{Aleksandrs Belovs\thanks{CSAIL, Massachusetts Institute of Technology, abelov@csail.mit.edu}}
\date{}

\begin{document}

\maketitle

\begin{abstract}
In this paper, we study the following variant of the junta learning problem.  We are given oracle access to a Boolean function $f$ on $n$ variables that only depends on $k$ variables, and, when restricted to them, equals some predefined function $h$.  The task is to identify the variables the function depends on.  
When $h$ is the XOR or the OR function, this gives a restricted variant of the Bernstein-Vazirani or the combinatorial group testing problem, respectively.

We analyse the general case using the adversary bound, and give an alternative formulation for the quantum query complexity of this problem.  We construct optimal quantum query algorithms for the cases when $h$ is the OR function (complexity is $\Theta(\sqrt{k})$) or the exact-half function (complexity is $\Theta(k^{1/4})$).  The first algorithm resolves an open problem from~\cite{montanaro:groupTesting}.  
For the case when $h$ is the majority function, we prove an upper bound of $O(k^{1/4})$.
All these algorithms can be made exact.

We obtain a quartic improvement when compared to the randomised complexity (if $h$ is the exact-half or the majority function), and a quadratic one when compared to the non-adaptive quantum complexity (for all functions considered in the paper).
\end{abstract}

\mycommand{atici}{At{\i}c{\i}\xspace}
\mycommand{hg}{\hat\gamma}
\section{Introduction}
Learning theory studies the problem of reconstructing functions from their values in various points.  In this paper, we study the problem of exact learning from membership queries.  In this problem, one is given oracle (black-box) access to a function $f\colon \{0,1\}^n\to\{0,1\}$ belonging to some fixed class of functions $\cC$ (usually called {\em concept class}).  The task is to identify the function using the smallest possible number of queries to the oracle.  
It is required to give the exact description of the function, not an approximation (although, it is allowed to err with small probability like $1/3$).

This is a broad area of research both classically and quantumly.  We shall highlight some of the results.  Classically, the problem was defined by Angluin~\cite{angluin:conceptLearning}.  Bshouty \etal~\cite{bshouty:oraclesForExactLearning} obtained upper and lower bounds on the randomised query complexity of learning a concept class $\cC$ exactly using a combinatorial parameter $0<\hg^{\cC}\le 1$ of the class.  More specifically, the query complexity is $O\sA[\frac{\log |\cC|}{\hg^\cC}]$ and $\Omega\sA[\frac1{\hg^\cC} + \log|\cC|]$.  

Quantumly, this problem was analysed (under the name of quantum oracle interrogation or identification) by van Dam~\cite{vanDam:oracleInterrogation} and Ambainis \etal~\cite{ambainis:oracleIdentification}.  Van Dam considered the case when $\cC$ consists of all Boolean functions on $n$ variables, where $n/2+O(\sqrt{n})$ quantum queries suffice, in contrast to $n$ queries required classically.  Ambainis \etal constructed a quantum $O(\sqrt{n\log |\cC|\log n}\log\log |\cC|)$-query algorithm for the general case.  
Finally, Kothari~\cite{kothari:oracleIdentification} gave a complete characterization of the quantum query complexity of this problem in terms of $n$ and $|\cC|$.

Servedio and Gortler~\cite{servedio:equivalencesQuantum} proved some quantum analogues of the results in~\cite{bshouty:oraclesForExactLearning}.  In particular, they showed that, for any concept class $\cC$, the quantum query complexity of learning $\cC$ exactly is $\Omega\sB[\frac1{\sqrt{\hg^\cC}} + \frac{\log|\cC|}n]$.  Using this result, they obtained that the deterministic complexity of the same problem is $O(nQ^3)$ where $Q$ is its quantum query complexity.  \atici and Servedio~\cite{atici:improvedBoundsLearning} constructed a quantum $O\sB[\frac{\log|\cC|\log\log|\cC|}{\sqrt{\hg^{\cC}}}]$-query algorithm for the same problem.

\paragraph{The problem and related work}
In this paper, we study the following learning problem proposed by Ambainis and Montanaro~\cite{montanaro:groupTesting}.
Let $h\colon\{0,1\}^k\to\{0,1\}$ be a fixed symmetric Boolean function.  
We are given oracle access to a Boolean function $f$ on $n\gg k$ variables that satisfies the following properties.  The function $f$ only depends on a subset $A$ of $k$ input variables, and, when restricted to these variables, the function equals $h$.  Thus, the learning problem reduces to identifying the set $A$.  

Functions that only depend on a small number of the input variables are called {\em juntas}.  Thus, our problem is related to the problem of learning and testing juntas, which has been studied both classically (see~\cite{blais:testingJuntas} and the references therein) and quantumly~\cite{atici:testingJuntas}.  
Note, however, that our settings are different from that of usual junta learning.  First, we have an additional promise that the function $f$ equals function $h$.  Second, we are allowed adaptive membership queries, not only samples.  And third, we have to find the function $f$ exactly, not an approximation.
The last two aspects make our settings different from the quantum PAC model~\cite{bshouty:learningDNF}.

A simple information-theoretical argument shows that $\Omega(\log|\cC|) = \Omega(k\log\frac nk)$ randomised queries are required to solve this problem classically.  Quantumly, as usual, one can do better.
One of the pioneering quantum algorithms, the Bernstein-Vazirani algorithm~\cite{bernstein:quantumComplexity}, can be stated in these settings.  The algorithm solves our problem for the case when $h$ is the XOR function.  It does so in one query, without an error, and, moreover, for all values of $k$ simultaneously.

Another example is the combinatorial group testing problem (despite the name, it is a {\em learning} problem).  
In this problem, a set $X$ of $n$ elements is given, and it is known that at most $k$ of them are marked.
For any subset $S\subseteq X$, it is possible to detect, in one query, whether $S$ contains a marked element.
The task is to identify all marked elements making as few queries as possible.
It corresponds to the case when $h$ is the OR function (if we additionally require having {\em exactly} $k$ marked elements).  This is a well-studied problem classically~\cite{du:combinatorialGroupTesting}.
Ambainis and Montanaro~\cite{montanaro:groupTesting} studied the quantum complexity of this problem and its special case, search with wildcards, that we do not define here.  The search with wildcards problem was resolved, but the complexity of the combinatorial group testing problem was only stated to lie between $\Omega(\sqrt{k})$ and $O(k)$.

The quantum counterfeit coin problem studied by Iwama \etal~\cite{iwama:quantumCounterfeit} is also closely connected to our work.  In this problem, one is given $n$ coins, and it is known that exactly $k$ of them are counterfeit.  All genuine coins have the same weight, all counterfeit coins have the same weight, and the counterfeit coins are strictly lighter than the genuine ones.  One is also given perfect scales, and the task is to find all counterfeit coins using as few weighing operations as possible.  More formally, the oracle accepts two disjoint equal-sized subsets $S,T\subseteq [n]$ as its input.  It replies with 0 if $S$ and $T$ contain equal number of counterfeit coins, and with 1 otherwise.  (I.e., one only gets to know whether the scales are balanced or not.)  Iwama \etal constructed a quantum algorithm that solves this problem in $O(k^{1/4})$ queries to the oracle.  No general lower bound is known for this problem.

%

\paragraph{Our contribution} 
In this paper, we do the following.  In \rf(sec:groupTesting), we resolve the question posed by Ambainis and Montanaro by describing a tight quantum $O(\sqrt{k})$-query algorithm for the combinatorial group testing problem (in its full generality, i.e., allowing less than $k$ marked elements).
In \rf(sec:representations), we use the adversary bound and representation theory to formulate an optimization problem for the quantum query complexity of our learning problem for any symmetric function $h$.  In \rf(sec:exactHalf), we solve this optimization problem when $h$ is the exact-half function (the function that evaluates to 1 iff exactly $\lfloor k/2\rfloor $ of the input variables equal 1).  The quantum query complexity of the learning problem turns out to be $\Theta(k^{1/4})$.  In \rf(sec:majority), we describe some partial results for the case when $h$ is the majority function.  Finally, in \rf(sec:nonadaptive), we show that most of the above algorithms can be made exact without increase in their complexity, and prove some no-go results for non-adaptive quantum algorithms.

\paragraph{Previous techniques}
Before discussing our techniques, let us describe some previously used techniques.  
One possibility is to apply the Grover search (as in the papers by Ambainis \etal~\cite{ambainis:oracleIdentification}, and \atici and Servedio~\cite{atici:improvedBoundsLearning}).  This gives at most quadratic speed-up.

Most of the papers, however, use the following {\em prepare-and-measure} strategy: A quantum state $\ket|\psi>$ is prepared, a tensor power $O_x^{\otimes T}$ of the input oracle is applied to the state, and the result is measured.  This strategy usually comes in one of the two variations.  The first one is {\em Fourier sampling}.  In this case, $T=1$ and $\ket|\psi>$ is the uniform superposition.  The resulting state, $O_x\ket|\psi>$, is measured in the Fourier basis.  This procedure is repeated many times, and when enough samples have been collected, they are processed by a classical subroutine to reconstruct $f$.  Notable examples are the DNF learning algorithm by Bshouty and Jackson~\cite{bshouty:learningDNF} and the junta learning algorithm by \atici and Servedio~\cite{atici:testingJuntas}, where this approach is mentioned explicitly (under the name of quantum example oracle in the first paper, and Fourier sampling oracle in the second one).

A more general variant is to show that the states $O_x^{\otimes T}\ket|\psi>$ and $O_y^{\otimes T}\ket|\psi>$ are almost orthogonal for all $x\ne y$, and then apply the  Pretty Good Measurement~\cite{hausladen:PGM} to distinguish them.  Examples here are Ref.~\cite{childs:hiddenShift, ettinger:hspQuery}.

Either way, the prepare-and-measure strategy usually can be made non-adaptive (see \rf(sec:prelim) for the definition).  This is a limitation.  For example, Zalka~\cite{zalka:GroverOptimal} showed that a non-adaptive quantum algorithm requires $\Omega(n)$ queries to solve the OR function, in contrast to the Grover search.  Childs \etal in~\cite{childs:hiddenShift} explain why their hidden shift algorithm performs sub-optimally on the delta function using this argument.

All these approaches are unsatisfactory for our problem.  First, the general results mentioned in the beginning of this section are useless here, because the quantum query complexity of our problems is less than $k$, which is much less than $n$ or $\log|\cC|$.  Next, we attain  super-quadratic speed-ups over randomised algorithms that is not possible by only using the Grover search. Finally, in \rf(sec:nonadaptive), we show that any {\em non-adaptive} quantum algorithm requires quadratically more queries than our algorithms.  This does not completely rule out the prepare-and-measure strategy, but shows that its easiest and most common one-shot variant does not work here.

It is also interesting to compare our algorithm for the exactly-half function to the algorithm for the counterfeit coin problem by Iwama \etal~\cite{iwama:quantumCounterfeit}.  After all, both algorithms attain complexity $O(k^{1/4})$, which is a quartic improvement to the randomised complexity.
We are not aware of any reduction in either of two directions.
Iwama \etal reduce the counterfeit problem to the Bernstein-Vazirani problem.  Indeed, if an even-sized subset $S$ contains even number of counterfeit coins, there exist dissections of $S$ into two equal-sized subsets having equal number of counterfeit coins.  These dissections can be detected using quantum amplitude amplification~\cite{brassard:amplification}.  It seems unlikely that a similar approach can be applied for the exact-half function.

\paragraph{Our techniques}
Instead of these techniques, we use the dual adversary bound.  The adversary bound is a lower bound on quantum query complexity first developed by Ambainis~\cite{ambainis:adv} in the form that is now known as the positive-weighted adversary.  Later, it was strengthened by H\o yer \etal~\cite{hoyer:advNegative} to the negative-weighted, or general adversary bound.  Reichardt \etal proved that this lower bound is tight by showing how the dual to the adversary bound can be converted into a quantum query algorithm~\cite{reichardt:spanPrograms, lee:stateConversion}.  Their algorithm is based on quantum walks.

Thus, a quantum query algorithm can be constructed by coming up with a feasible solution to the dual adversary bound.  There has been some work in this vein.  One example is provided by algorithms for formulae evaluation~\cite{reichardt:formulae, zhan:treesWithHiddenStructure}.
Another line of development is learning graphs~\cite{belovs:learning}.  They were applied to improve quantum query complexity of triangle and other subgraph detection~\cite{lee:learningTriangle, belovs:learningClaws}, and the $k$-distinctness problem~\cite{belovs:learningKDist}.  In general, learning graphs work well for Boolean functions with small 1-certificates.  
Clearly, both of these general approaches do not work here.  Indeed, our problem does not have a nice formula description, nor does it have Boolean output, nor small certificates.

Instead of that, we construct a feasible solution to the dual adversary from scratch.
Let us give a short overview of our construction.  For precise formulations of the adversary bound, the reader may refer to \rf(sec:prelim).  
Informally, the dual adversary bound~\rf(eqn:advDual) boils down to distinguishing inputs $A,B\in\cC$ using queries~\rf(eqn:advDualCondition).  
In the following informal exposition, we analyse complexity of distinguishing $A$ and $B$ using both a usual randomised algorithm and the the adversary bound, and compare the two.
Although obtaining equality in~\rf(eqn:advDualCondition), and not a lower bound like in~\rf(eqn:advPositive), is important, we ignore this issue for now.

We start with combinatorial group testing, which corresponds to the case when $h$ is the OR function.
Assume we want to distinguish $k$-subsets $A,B\subseteq[n]$.  Moreover, we want to do so regardless of the distance $\ell = |B\setminus A|$.
A simple strategy is to take a subset $S\subseteq[n]$ by including each element of $[n]$ with probability $p$ independently at random, and hope that exactly one of $S\cap A$ and $S\cap B$ is empty.

Classically, the worst case is when the distance $\ell = 1$.  
In this case, conditioned on $S\cap A=\emptyset$, the probability that $S$ distinguishes $A$ and $B$ (i.e., that $S\cap B\ne\emptyset$) is $p$.
But taking $p\gg 1/k$ does not make much sense, because then the probability that $S$ does not intersect $A$ is too small.

The dual adversary, however, allows for additional tricks.
In particular, we may ``condition'' on $S$ and $A$ having intersection of size at most 1.  That is, the queries $S$ with $|S\cap A|\ge 1$ count neither towards the complexity, nor towards distinguishing $A$ and $B$.  (The same, clearly, applies for $B$ as well.)
Thus, in this settings, we may even take $p=1/2$, which increases the chances of $A$ and $B$ being distinguished.

But when $\ell$ is, say, $k$, the choice of $p=1/2$ does not work.  Indeed, conditioned on $A\cap S = \emptyset$, the probability of $|S\cap B| = 1$ is very small.  (Remember, we do not use $S$ for $B$ if $|S\cap B|>1$.)  In this case, $p=1/k$ is a much better choice.  In the final solution, we take $p\in(0,1)$ uniformly at random that, a bit surprisingly, works for all values of $\ell$.

Thus, our solution to the combinatorial group testing problem is somewhat {\em ad hoc}.
The analysis is so simple because we may assume that $S$ intersects $A$ in either 0 or 1 element.
If $h$ is the majority function, it is suboptimal to condition that $|S\cap A|$ is $\lceil k/2 \rceil-1$ or   $\lceil k/2 \rceil$.  Indeed, assume $A\cap B = \emptyset$.  Then, regardless of $A\cap S$, the probability is at most $O(1/\sqrt{k})$ that $|S\cap B| \in \{\lceil k/2 \rceil-1,  \lceil k/2 \rceil \}$.  Thus, to solve this case, we would have to take other intersection sizes as well, and that would make the analysis much more complicated.

Instead of sticking to this {\em ad hoc} solution, we use an approach that is guaranteed to be tight.
Without loss of generality, we may assume that the optimal solution $\Gamma$ to the adversary lower bound~\rf(eqn:advPrimal) is symmetric with respect to permuting the elements of $[n]$.  Then, the matrix $\Gamma$ can be uniquely described by $k+1$ real numbers.  We use representation theory of the symmetric group and obtain necessary and sufficient conditions that these numbers must satisfy.  A feasible solution to the dual problem again gives a quantum query algorithm.

Unfortunately, the resulting optimization problem is still very complicated.  We were able to obtain a feasible solution, when $h$ is the majority or the exact-half function, using that these functions are symmetric about the weight $k/2$.  But applying these scheme for the OR function, for instance, would be much more complicated than our previous {\em ad hoc} solution.  Our solutions for majority and exact-half are essentially equivalent, but for exact-half, the solution turns out to be tight.  Generalizing this solution to the exact-$\ell$ or the $\ell$-threshold function is an open problem.

\section{Preliminaries}
\label{sec:prelim}
We use $[n]$ to denote the set $\{1,2,\dots,n\}$, and $2^A$ to denote the set of subsets of $A$.  A $k$-subset is a subset of size $k$.

All matrices in the paper have real entries.  $A^*$ denotes the adjoint (transposed) matrix of $A$.
If $A$ is a matrix, by $A\elem[i,j]$, we denote the element on the intersection of row $i$ and column $j$.  
By $\|A\|$ we denote the spectral norm of $A$ (the maximal singular value), and by $\normtr|A|$ we denote the trace norm of $A$ (the sum of the singular values).  By $\ip<A,B>$ we denote the inner product between the matrices: $\ip<A,B> = \tr(A^*B)$.

We assume familiarity with basic probability theory, and we repeatedly use the following well-known result about binomial coefficients:
\begin{lem}
\label{lem:binomial}
If $n$ and $k$ are positive integers satisfying $k=O(\sqrt{n})$, then ${n\choose \floor[n/2]\pm k} = \Theta(2^n/\sqrt{n})$.
\end{lem}

\paragraph{Quantum query complexity}
Now we define quantum query complexity both in its standard and non-adaptive variants.  For a more complete treatment refer to~\cite{buhrman:querySurvey} for query complexity and~\cite{montanaro:nonadaptive} for non-adaptive query complexity.
A quantum query algorithm is defined as a sequence of unitary transformations alternated with the oracle calls:
\begin{equation}
\label{eqn:queryAlgDef}
U_0\to O_x\to U_1\to O_x \to \cdots \to U_{T-1} \to O_x\to U_T.
\end{equation}
Here $U_i$s are arbitrary unitary transformations independent of the input.  The oracle $O_x$ is the same in all places, and it depends on the input string $x=(x_i)$ as $\ket i|i>\ket v|b>\mapsto \ket i|i>\ket v|b+x_i>$ where the addition is performed modulo 2.  Other registers besides $\reg i$ and $\reg v$ are left intact.  The computation starts in a predefined state $\ket|0>$.  After all the operations in~\rf(eqn:queryAlgDef) are performed, some predefined output register is measured.  We say that the algorithm evaluates a function $f$ if, for any $x$ in the domain, the result of the measurement is $f(x)$ with probability at least $2/3$.  The number $T$ is the {\em query complexity} of the algorithm.  The smallest value of $T$ among all algorithms evaluating $f$ is the quantum query complexity of $f$, and is denoted by $Q(f)$.

Thus, we see that a quantum algorithm can prepare the input to the next oracle query depending on the results of the previous oracle calls.  In many cases, this is crucial for obtaining a good algorithm.  But, in some cases, the input to the oracle does not depend on the output of its previous executions.  This is captured by the notion of {\em non-adaptive} quantum query complexity.  In such an algorithm, we assume that all the oracle calls happen simultaneously in parallel.  More formally, a non-adaptive quantum query algorithm is of the form $U_0\to O_x^{\otimes T}\to U_1$.  The non-adaptive quantum query complexity of $f$ is then defined similarly to the adaptive case.

\paragraph{Formulation of the problem}
Let us rigorously define our version of the learning problem.   
Let $h\colon\{0,1\}^k\to\{0,1\}$ be a symmetric Boolean function.  
It is uniquely defined by a subset $W_h\subseteq \{0,\dots,k\}$ such that $h(x)=1$ iff $|x|\in W_h$, where $|x|$ stands for the Hamming weight of $x$.
Let $n\ge k$ be a positive integer, and $\cC$ denote the set of all $k$-subsets of $[n]$.
If $A\in\cC$, we define the function $f_A:\{0,1\}^n\to\{0,1\}$ by $f_A(x) = h(x_A)$ where $x_A$ is the restriction of the input string $x$ to the positions in $A$.  It is more convenient to identify the input string $x$ with the subset $S\subseteq[n]$ defined by $i\in S$ iff $x_i=1$.  Thus, $f_A(S)=1$ iff $|A\cap S|\in W_h$.

The learning problem $L_h^n\colon \{0,1\}^{\{0,1\}^n}\to 2^{[n]}$ is defined by $L_h^n(f_A) = A$.  
Thus, $h$ is fixed and known to the learner in advance, the inputs are the functions $f_A$ (which can be identified with the elements of $\cC$), and the input variables are the input strings to $f_A$ (which can be identified with the subsets of $[n]$).

It is easy to see that the quantum query complexity $Q(L_h^n)$ is a non-decreasing function in $n$.  There also exists an upper bound on $Q(L_h^n)$ independent of $n$.  For instance, one may take the complexity of the Fourier sampling algorithm like in~\cite{atici:testingJuntas}, since its behaviour does not depend on $n$.  Hence, there exists $\lim_{n\to\infty} Q(L_h^n)$, which we denote by $Q(L_h)$, and which we are mostly interested in.

\paragraph{Adversary Bound}
Next, we define the adversary bound tailored to our special case of $L_h^n$.  An adversary matrix $\Gamma$ is a $\cC\times\cC$ real symmetric matrix with zeroes along the diagonal.  Introducing an abuse of notation, let $\Gamma\circ\Delta_S$ denote the submatrix of $\Gamma$ formed by the rows in $\{A\in\cC\mid f_A(S)=0\}$ and the columns in $\{B\in\cC\mid f_B(S)=1\}$.

The adversary bound $\Adv(L_h^n)$ is equal to the (common) optimal value of the following two optimisation problems:
\begin{subequations}
\label{eqn:advPrimal}
\begin{alignat}{3}
&\mbox{\rm maximise} &\quad& \|\Gamma\| \label{eqn:advPrimalObjective}\\
& \mbox{\rm subject to}&& \|\Gamma\circ\Delta_S\|\le 1 &\quad& \text{\rm for all $S\subseteq[n]$;} \label{eqn:advPrimalCondition}\\
&&& \Gamma\elem[A,A] = 0 && \mbox{\rm for all $A\in\cC$.} \label{eqn:advPrimalDiagonal}
\end{alignat}
\end{subequations}
and
\begin{subequations}
\label{eqn:advDual}
\begin{alignat}{3}
&\mbox{\rm minimise} &\quad& \max_{A\in \cC}\sum\nolimits_{S\subseteq[n]} X_S\elem[A,A]  \label{eqn:advDualObjective} \\
& \mbox{\rm subject to}&& \sum\nolimits_{S\colon f_A(S)\ne f_B(S)} X_S\elem[A, B] = 1 &\quad& \text{\rm for all $A\ne B$ in $\cC$;} \label{eqn:advDualCondition} \\
&&& X_S\succeq 0 && \mbox{\rm for all $S\subseteq[n]$,} \label{eqn:advDualSemidefinite}
\end{alignat}
where $X_S$ are $\cC\times\cC$ positive semi-definite matrices (see~\cite[Theorem 6.2]{reichardt:spanPrograms} for the proof of the equality of both problems).  The adversary bound is very useful because of the following result:
\begin{thm}[\cite{hoyer:advNegative, lee:stateConversion}]
The quantum query complexity of a function $f$ equals $\Theta(\Adv(f))$.
\end{thm}

Using this theorem, we can estimate $\Adv(L_h)$ instead of $Q(L_h)$.  Here we denote $\Adv(L_h) = \lim_{n\to\infty} \Adv(L_h^n)$.  The limit exists because $\Adv(L_h^n)$ is a non-decreasing function in $n$.

An important special case of the adversary bound is the positive-weighted adversary, which we denote by $\pAdv(L_h^n)$.  It is a slight modification of the original version by Ambainis~\cite{ambainis:adv}.  It is strictly weaker than the general bound, but it is usually much easier to apply.
The positive-weighted adversary is defined as in~\rf(eqn:advPrimal) and~\rf(eqn:advDual) with the following modifications.  In~\rf(eqn:advPrimal), we require all the entries of $\Gamma$ to be non-negative.
In~\rf(eqn:advDual), we replace condition~\rf(eqn:advDualCondition) by the following one~\cite[Eq. (3.7)]{spalek:advEquivalent}:
\begin{equation}
\label{eqn:advPositive}
\qquad \sum\nolimits_{S\colon f_A(S)\ne f_B(S)} X_S\elem[A, B] \ge 1 \qquad \text{\rm for all $A\ne B$ in $\cC$;}
\end{equation}
\end{subequations}

\section{Combinatorial Group Testing}
\label{sec:groupTesting}
In this section, we describe a quantum query algorithm for the combinatorial group testing problem.  We solve the problem in its original form, which deviates slightly from our version of the learning problem.
Let us reformulate the problem.  Let $k<n$ be fixed positive integers, and $\cC$ consist of all subsets of $[n]$ of sizes at most $k$.  For each $A\in\cC$, the function $f_A\colon 2^{[n]}\to \{0,1\}$ is defined by
\[
f_A(S) = 
\begin{cases}
1,& \text{if $A\cap S \ne\emptyset$;} \\
0,& \mbox{otherwise.}
\end{cases}
\]
We are given oracle access to $f_A$, and the task is to detect $A$.  The difference with the $L_{\text{OR}}$ problem is that we allow $A$ of size less than $k$.  In this section, we prove the following result:

\begin{thm}
\label{thm:combGroupTest}
The quantum query complexity of the combinatorial group testing problem is $\Theta(\sqrt{k})$.
\end{thm}

The lower bound can be proved by a reduction from the unordered search, refer to~\cite{montanaro:groupTesting} for more detail.  Here we prove the upper bound.
We do so by constructing a feasible solution to~\rf(eqn:advDual).  
This is done in two steps:  First, we define rank-1 matrices $Y_S(p)$, and then build the matrices $X_S$ from them.

\mycommand{btr}{\mathop{\!\triangle}}
Let $P$ be the binomial probability distribution on $[n]$ with probability $p$.  Recall that it is a probability distribution on the subsets of $[n]$, where each element of $[n]$ is included into the subset independently with probability $p$.  By $P(S)$, we denote the probability of sampling $S$ from $P$: $P(S) = p^{|S|}(1-p)^{n-|S|}$. Finally, let $\btr$ denote the symmetric difference of sets.

We define $Y(p) = (Y_S(p))_{S\subseteq[n]}$ by
\[
Y_S(p) = \frac{ P(S)} {2 p}\; \psi \psi^* \succeq 0,
\]
where
\[
\psi\elem[A] =
\frac{1}{(1-p)^{|A|/2}} \times
\begin{cases}
\displaystyle \sqrt[4]{k p/ (1-p)}, & \text{if $|A\cap S| = 0$;}\\
\displaystyle \sqrt[4]{(1-p)/(kp)}, & \text{if $|A\cap S| = 1$;}\\
0,& \mbox{otherwise;}
\end{cases}
\]
for all $A\in\cC$.
In this notation, 
\begin{align*}
\sum_{S\subseteq[n]} Y_S(p)\elem[A,A] &= 
\frac1{2 p\, (1-p)^{|A|}}\sC[{ \pr_{S\sim P}\skA[|S\cap A| = 0] \sqrt{\frac{kp}{1-p}} \;+\; \pr_{S\sim P}\skA[|S\cap A| = 1] \sqrt{\frac{1-p}{kp}}\; }] \\
&= \frac1{2 p\, (1-p)^{|A|}}\sC[{ (1-p)^{|A|} \sqrt{\frac{kp}{1-p}} \;+\; |A| p\,(1-p)^{|A|-1} \sqrt{\frac{1-p}{kp}}\; }] \le \sqrt{\frac{k}{p\,(1-p)}}\;.
\end{align*}

Now we fix two distinct elements $A,B$ of $\cC$.
An element $A$ is used in $Y_S$ only if $|S\cap A|\le 1$.  Thus, we are only interested in $S\subseteq [n]$ such that $|A\cap S|+|B\cap S|=1$.  Thus,
\begin{align*}
\sum_{S\colon f_A(S)\ne f_B(S)} Y_S(p)\elem[A, B] &= \frac{\Pr_{S\sim P} \skA[|A\cap S|+|B\cap S|=1]}{2p\;(1-p)^{(|A|+|B|)/2}}\\
&= \frac{|A\btr B|\, p\, (1-p)^{|A\cup B|-1}}{2p\;(1-p)^{(|A|+|B|)/2}}  = \frac{|A\btr B|}{2} (1-p)^{\frac{|A\btr B|}2 - 1}\;.
\end{align*}

Now, for each $S\subseteq[n]$, let
\[
X_S = \int_0^{1} Y_S(p)\; \dd p\;.
\]
First, each $X_S$ is positive semi-definite, because positive semi-definite matrices form a convex cone.  Next, for any $A\in\cC$:
\[
\sum_{S\subseteq[n]} X_S\elem[A,A] \le \sqrt{k} \int_0^1 \frac{\dd p}{\sqrt{p\,(1-p)}} = \pi\sqrt{k}\;.
\]
And finally, for all $A\ne B$ in $\cC$:
\[
\sum_{S\colon f_A(S)\ne f_B(S)} X_S\elem[A, B] = \frac{|A\btr B|}{2} \int_0^{1} (1-p)^{\frac{|A\btr B|}2 - 1} \; \dd p = 1.
\]

\def\S{\mathbb{S}}
\def\up{{\uparrow}}
\def\down{{\downarrow}}
\def\tr{\mathop{\mathrm{tr}}}

\section{Application of Representation Theory}
\label{sec:representations}
In the previous section, we described an {\em ad hoc} construction of a feasible solution to~\rf(eqn:advDual) when $h$ is the OR function.
In this section, we use representation theory to give an alternative description for $\Adv(L_h)$ that works for any function $h$.
We work with the lower bound~\rf(eqn:advPrimal), because it has a very simple structure.
In the next two sections, we use duality to the new formulation to prove that the quantum query complexity of the $L_{\text{EXACT-HALF}_k}$ and the $L_{\text{MAJORITY}_k}$ problems is $O(k^{1/4})$.

Let $h:\{0,1\}^k\to\{0,1\}$ be a symmetric function defined by the subset $W_h$ of weights, i.e., $h(x)=1$ iff $|x|\in W_h$.  The search for an adversary matrix for the function $L_h$ turns out to be equivalent to the search for a list of real numbers $d=(d_0,\dots,d_k)$ satisfying the constraints we are about to describe.

Let $m\le k$ be a positive integer and $0<p<1$ be a real number.  
We make use of Krawtchouk polynomials for probability $p$.  These polynomials are orthogonal with respect to the binomial distribution (see~\cite{szego:orthogonal} for the general definition, and~\cite{krasikov:KrawtchoukSurvey} for the special case $p=1/2$, which we use in Sections~\ref{sec:exactHalf} and~\ref{sec:majority}).
We treat them as column vectors in $\bR^{m+1}$ and also include the weight (due to the weight, they cease to be polynomials).  With this modification, the definition is as follows:
\begin{equation}
\label{eqn:Kdef}
K^{(m,p)}_t\elem[x] = \sqrt{{m\choose x}p^x(1-p)^{m-x}}\; \sum_{i=0}^t (-1)^i p^{t-i}(1-p)^i {x\choose i}{m-x\choose t-i},
\end{equation}
where $t,x\in\{0,\dots,m\}$.  Let $\varkappa^{(m,p)}_t = K^{(m,p)}_t/\|K^{(m,p)}_t\|$ be the corresponding normalised vectors.  Thus, $\{\varkappa_t^{(m,p)}\}$, for fixed $m$ and $p$, form an orthonormal basis of $\bR^{m+1}$.  We use the list $d$ to define the matrices
\begin{equation}
\label{eqn:Mdef}
M_{m, p}^{(d)} = \sum_{i=0}^{m} d_{k-i} \varkappa^{(m,p)}_{m-i}\sA[\varkappa_{m-i}^{(m,p)}]^*.
\end{equation}

Let $0\le t\le k-m$ be an integer, and define $W_1(t) = \{\ell \in\bZ\mid 0\le \ell\le m,\; \ell+t\in W_h \}$, and $W_0(t) = \{0,\dots,m\}\setminus W_1(t)$.  Let
\begin{equation}
\label{eqn:Mdef2}
M_{m,p,t}^{(d)} = M_{m,p}^{(d)}\elem[W_0(t), W_1(t)]
\end{equation}
be the submatrix of $M_{m,p}^{(d)}$ formed by the rows in $W_0(t)$ and the columns in $W_1(t)$.

The aim of this section is to prove the following result:
\begin{thm}
\label{thm:main}
For any symmetric function $h$, $\Adv(L_h)$ equals the supremum of $\max_i d_i$ over all lists of real numbers $d=(d_0,\dots,d_k)$ satisfying the following constraints:
\itemstart
\item $d_k=0$, and 
\item for all integers $0<m\le k$, $0\le t\le k-m$, and reals $0<p<1$, we have
$\|M_{m,p,t}^{(d)} \| \le 1$, where $M_{m,p,t}^{(d)}$ is defined in~\rf(eqn:Mdef2).
\itemend
\end{thm}

In order to prove this theorem, we need some basic results from representation theory of the symmetric group.
These results are only used in this section.
The reader may refer to a textbook on the topic like, e.g.,~\cite{sagan:symmetricGroup}, or to the appendix, where we briefly formulate the required notions and results.

If $N$ is a finite set, let us denote by $\S_N$ the symmetric group on $N$.  We consider modules over the group algebra $\bR G$ where $G$ is either a symmetric group or a direct product of two symmetric groups.

Fix an integer $n$, and consider the problem $L_h^n$.  Let also $N=[n]$.
The rows and the columns of an adversary matrix $\Gamma$ are labelled by $k$-subsets of $N$.  
The problem is symmetric with respect to the permutations of variables, so by~\cite{hoyer:advNegative} we may assume that $\Gamma$ is symmetric with respect to $\S_{N}$.  More specifically, $\Gamma$ does not change if we simultaneously transform the labels of its rows and columns by $\{a_1,\dots,a_k\}\mapsto \{\pi a_1,\dots,\pi a_k\}$ for some $\pi\in\S_N$.  

The real vector space with the set of $k$-subsets of $N$ as its orthonormal basis, and the above action of $\bS_N$, is the permutation $\bR\S_N$-module corresponding to the partition $(n-k,k)$ of $n$.  We denote it by $M(N,k)$.  We denote the basis element of $M(N,k)$ corresponding to $A$ by $A$ itself.

\mycommand{subgroup}{\bS_{N_0}\times \bS_{N_1}}
\mycommand{subalgebra}{\bR(\subgroup)}
Now consider $\|\Gamma\circ \Delta_S \|$ for $S\subseteq N$.  
We denote $N_0 = N\setminus S$, $N_1 = S$, $n_0=|N_0|$, and $n_1=|N_1|$.
Then, $\Gamma\circ\Delta_S$ is symmetric with respect to $\S_{N_0}\times \S_{N_1}$.  Thus, we have to understand how the $\bR\S_N$-module $M(N,k)$ behaves under restriction to this subgroup.
It is easy to see that
\begin{equation}
\label{eqn:Mrestricted}
M(N,k) \down_{\S_{N_0}\times \S_{N_1}} = \bigoplus_{k_0+k_1=k} M(N_0, k_0)\otimes M(N_1, k_1),
\end{equation}
where $A\otimes B$, with $A$ being a basis element of $M(N_0,k_0)$ and $B$ being a basis element of $M(N_1,k_1)$, is understood as the basis element $A\cup B$ of $M(N,k)$.
We continue using the convention that $A\otimes B$ is the disjoint union of $A$ and $B$ later, for instance, in~\rf(eqn:vkDef).

Let $\Pi_1$ be the projector onto the spaces on the right-hand side of~\rf(eqn:Mrestricted) with $k_1\in W_h$, and $\Pi_0$ be the projector onto the orthogonal complement of this space.  Then, 
\begin{equation}
\label{eqn:GammaDelta}
\|\Gamma\circ\Delta_S\| = \|\Pi_0\Gamma\Pi_1\|.
\end{equation}

The following result describes the decomposition of $M(N,k)$ into irreducible submodules.  They are isomorphic to the Specht modules $S(N,t)$ corresponding to partitions $(n-t,t)$ of $n$.  The modules with different values of $t$ are not isomorphic.  
The lemma follows from general theory~\cite[Sections 2.9 and 2.10]{sagan:symmetricGroup}.  We give a proof in the appendix.

\begin{lem}
\label{lem:specht}
The $\bR\S_N$-module $M(N,k)$ has the following decomposition into irreducible submodules:
\(
M(N,k) = \bigoplus_{t=0}^k S_k(N,t),
\)
where each $S_k(N,t)$ is isomorphic to $S(N,t)$.  The submodule $S_k(N,t)$ is spanned by the vectors
\begin{equation}
\label{eqn:vkDef}
v_k(N,t,a,b) = 
\sA[\{a_1\}-\{b_1\}]\otimes \cdots\otimes \sA[\{a_t\}-\{b_t\}] \otimes \sC[\sum_{A\subseteq N\setminus\{a_1,\dots,a_t,b_1,\dots,b_t\}\colon |A|=k-t} A ]
\end{equation}
defined by disjoint sequences $a=(a_1,\dots,a_t)$ and $b=(b_1,\dots,b_t)$ of pairwise distinct elements of $N$.
The dimension of $S(N,t)$ is ${n\choose t}-{n\choose t-1}$.

Moreover, the only (up to a scalar) $\bR\S_N$-isomorphism of $S_k(N,t)$ onto $S_\ell(N,t)$ maps the vector $v_k(N,t,a,b)$ into $v_{\ell}(N,t,a,b)$ for any choice of $a$ and $b$.
\end{lem}

We define $S_{k_0}(N_0, t_0)$ and $S_{k_1}(N_1,t_1)$ similarly.
By combining~\rf(eqn:Mrestricted) and \rf(lem:specht), we get that the irreducible $\bR(\S_{N_0}\times\S_{N_1})$-submodules of $M(N,k) \down_{\S_{N_0}\times \S_{N_1}}$ are $S_{k_0}(N_0,t_0)\otimes S_{k_1}(N_1,t_1)$, where
\begin{equation}
\label{eqn:tcondition}
k_0+k_1=k,\quad 0\le t_0\le k_0,\quad\mbox{and}\quad 0\le t_1\le k_1.
\end{equation}
Two submodules of this form are isomorphic iff their values of $t_0$ and $t_1$ are equal.  
Thus, the canonical submodules of $M(N,k) \down_{\S_{N_0}\times \S_{N_1}}$ are
\[
R(t_0,t_1) = \bigoplus_{\text{$k_0,k_1$ satisfy~\rf(eqn:tcondition)}} S_{k_0}(N_0,t_0)\otimes S_{k_1}(N_1,t_1),
\]
and the multiplicity of $S(N_0,t_0)\otimes S(N_1,t_1)$ in $M(N,k)$ is $k+1-t_0-t_1$.

By Schur's lemma, in a suitable basis of $R(t_0,t_1)$, any $\subalgebra$-homomorphism from $R(t_0,t_1)$ to itself is of the form $A\otimes I_{t_0,t_1}$, where $A$ is an $(k+1-t_0-t_1)\times(k+1-t_0-t_1)$ matrix, and $I_{t_0,t_1}$ is the identity matrix in $S(N_0,t_0)\otimes S(N_1,t_1)$.
For each $(t_0,t_1)$, we choose the basis $\{e_\ell\}_{\ell\in\{0,\dots,k-t_0-t_1\}}$ for the matrix $A$ so that $(e_\ell e_\ell^*)\otimes I_{t_0,t_1}$ projects onto $S_{k-t_1-\ell}(N_0,t_0)\otimes S_{t_1+\ell}(N_1,t_1)$.
With this choice of the basis, we have that
\begin{equation}
\label{eqn:ADelta}
\Pi_0 (A\otimes I_{t_0,t_1}) \Pi_1 = A\elem[W_0(t_1), W_1(t_1)]\otimes I_{t_0,t_1},
\end{equation}
where $\Pi_0$ and $\Pi_1$ are as in~\rf(eqn:GammaDelta), and $W_0$ and $W_1$ are as in~\rf(eqn:Mdef2).

Let $\Pi_k(N,t)$ denote the orthogonal projector onto $S_k(N,t)$.  Again, by Schur's lemma,
\[
\Pi_k(N,t) = \bigoplus_{t_0,t_1} A^{(t)}_{t_0,t_1}\otimes I_{t_0,t_1}
\]
for some matrices $A^{(t)}_{t_0,t_1}$.  By the Littlewood-Richardson rule~\cite[Section 4.9]{sagan:symmetricGroup}, 
\begin{equation}
\label{eqn:reductionDecompose}
S(N,t)\down_{\S_{N_0}\times \S_{N_1}} \cong \bigoplus_{t_0+t_1\le t} S(N_0,t_0)\otimes S(N_1,t_1),
\end{equation}
so $A_{t_0,t_1}^{(t)}$ is zero if $t<t_0+t_1$, and, otherwise, it is a rank-1 orthogonal projector (as the corresponding multiplicity is 1).

At the heart of the proof of \rf(thm:main) is the following observation (recall that the matrices $A_{t_0,t_1}^{(t)}$ depend on the values of $n$ and $n_1$):  

\begin{lem}
\label{lem:convergence}
For any $0<p<1$, the projector $A_{t_0,t_1}^{(t)}$ tends to the projector onto $\varkappa^{(k-t_0-t_1, p)}_{t-t_0-t_1}$ as $n\to\infty$ and $n_1/n\to p$.
Moreover, the convergence is uniform for $c<p<1-c$ where $c>0$ is any constant.

On the other hand, there exists a bound $\eps_c$ satisfying $\lim_{c\to0} \eps_c = 0$, such that
$\|A_{t_0,t_1}^{(t)} - e_{t-t_0-t_1}e_{t-t_0-t_1}^* \|\le \eps_c$ if $n_1$ is less than $cn$, and
$\|A_{t_0,t_1}^{(t)} - e_{k-t}e_{k-t}^* \|\le \eps_c$ if $n_1$ is more than $(1-c)n$.
\end{lem}

We prove the lemma at the end of the section.
For now, let us show how the lemma can be used to prove \rf(thm:main).

Assume that $\Adv(L_h)=Q$.  As noticed in \rf(sec:prelim), $Q<\infty$.  Then, for each $n$, let $\Gamma^{(n)}$ be an optimal solution to~\rf(eqn:advPrimal).  We may assume that $\|\Gamma^{(n)}\|$ is an eigenvalue of $\Gamma^{(n)}$, otherwise replacing $\Gamma^{(n)}$ by $-\Gamma^{(n)}$.
By Schur's lemma, we may also assume that 
\begin{equation}
\label{eqn:GammaN}
\Gamma^{(n)} = \sum_{t=0}^k d_t^{(n)} \Pi_k(N,t).
\end{equation}
Consider the vectors $d^{(n)} = (d_t^{(n)})$.  As the absolute values of all $d_t^{(n)}$ are bounded by $Q$, the Bolzano-Weierstrass theorem gives a convergent subsequence $d^{(n_1)}, d^{(n_2)},\dots$.  We define $d=(d_t)$ as the limit of this subsequence.  
Clearly, $\max_t d_t = Q$.

Next, $\tr \Pi_k(N,t) = {n\choose t}-{n\choose t-1}$.  Hence, $\tr \Pi_k(N,k)$ overwhelms the traces of all other projectors in~\rf(eqn:GammaN) as $n\to\infty$.  Thus, by~\rf(eqn:advPrimalDiagonal),
\begin{equation}
\label{eqn:dk}
d_k = \lim_{i\to\infty} d_k^{(n_i)} = \lim_{i\to\infty} \frac{\tr \Gamma^{(n_i)}}{{n_i\choose k}-{n_i\choose k-1}} = 0.
\end{equation}
This proves the first constraint in \rf(thm:main).  The second constraint follows from \rf(lem:convergence) and~\rf(eqn:ADelta).

Now assume $d$ is an optimal solution to the optimization problem in \rf(thm:main), and let $Q=\max_t d_t$.  We define $\Gamma^{(n)}$ as in~\rf(eqn:GammaN), where $d_t^{(n)} = d_t$ for $t<k$, and $d_k^{(n)}$ is chosen so that $\tr(\Gamma^{(n)})=0$.  Then, due to symmetry, all diagonal entries of $\Gamma^{(n)}$ are equal to zero.  Also, similarly to~\rf(eqn:dk), $\lim_{n\to\infty} d^{(n)}_k=0$.

Choose $c>0$ so that $\eps_c \le 1/(2(k+1)Q)$, where $\eps_c$ is as in \rf(lem:convergence).  If $|S|/n < c$ or $|S|/n > 1-c$, then $\|\Gamma^{(n)}\circ\Delta_S\|\le 1/2$ for {\em any} choice of $d$ satisfying $\max_t d_t \le Q$.  If $|S|/n\to p$ with $c<p<1-c$, then $\lim_{n\to\infty}\|\Gamma^{(n)}\circ\Delta_S\| = 1$ by \rf(lem:convergence) and~\rf(eqn:ADelta) again.

\pfstart[Proof of \rf(lem:convergence)]
Fix two sequences of pairwise distinct elements in $N_0$: $a=(a_1,\dots,a_{t_0})$ and $b=(b_1,\dots,b_{t_0})$, and two sequences $a'=(a'_1,\dots,a'_{t_1})$ and $b'=(b'_1,\dots,b'_{t_1})$ in $N_1$.  In order to find the vector onto which $A_{t_0,t_1}^{(t)}$ projects, it suffices to find a linear combination of the vectors 
\[
\sfig{v_{k_0}(N_0,t_0,a,b)\otimes v_{k_1}(N_1,t_1,a',b')\midA \text{$k_0$, $k_1$ satisfy~\rf(eqn:tcondition)}}
\]
that belongs to $S_k(N,t)$.  

Clearly, $(\{a_1\}-\{b_1\})\otimes\cdots\otimes  (\{a_{t_0}\}-\{b_{t_0}\})$ and 
$(\{a'_1\}-\{b'_1\})\otimes\cdots\otimes  (\{a'_{t_1}\}-\{b'_{t_1}\})$ factor out in any linear combination,
so we can remove the elements in $a,b,a'$ and $b'$, and consider the case $t_0=t_1=0$.  The removal has the effect that $t$ gets reduced by $t_0+t_1$, $k_0$ by $t_0$, $k_1$ by $t_1$, $n_0$ by $2t_0$, and $n_1$ by $2t_1$.  The effect on $t$, $k_0$ and $k_1$ is reflected in the statement of the lemma, and the change in $n_0$ and $n_1$ is not substantial, as we assume $n\to\infty$.

So, it suffices to consider the case $t_0=t_1=0$.  In this case, the vector
\begin{equation}
\label{eqn:vek1}
\frac1{t!} \sum_{a,b} v_k(N,t,a,b),
\end{equation}
where the sum is over all sequences $a$ in $N_0$ and $b$ in $N_1$, is a linear combination of the vectors $\sfig{v_{k_0}(N_0,0,\emptyset,\emptyset)\otimes v_{k_1}(N_1,0,\emptyset,\emptyset)\midA k_0+k_1=k}$.  More specifically, the coefficient of $v_{k_0}(N_0,0,\emptyset,\emptyset)\otimes v_{k_1}(N_1,0,\emptyset,\emptyset)$ in~\rf(eqn:vek1) is
\begin{equation}
\label{eqn:coeff}
\sum_{i=0}^t (-1)^{i} {k_0\choose t-i}{k_1\choose i} (n_0-k_0)^{\underline{i}}\;(n_1-k_1)^{\underline{t-i}}\;,
\end{equation}
where $a^{\underline b}=a(a-1)\cdots(a-b+1)$ denotes the falling power.  That is, we claim that if $A$ is a $k_0$-subset of $N_0$ and $B$ is a $k_1$-subset of $N_1$, then the coefficient of $A\otimes B$ in~\rf(eqn:vek1) is~\rf(eqn:coeff).
Indeed, $i$ is the number of the elements of $B$ used in the sequence $b$, ${k_1\choose i}$ is the number of ways to choose them, $(n_0-k_0)^{\underline{i}}$ is the number of ways to choose the elements of $N_0\setminus A$ that serve as the corresponding element of the sequence $a$ (they do not appear in the product).  Calculations for $A$ are similar.

Taking the norm of the vector $v_{k_0}(N_0,0,\emptyset,\emptyset)\otimes v_{k_1}(N_1,0,\emptyset,\emptyset)$ into account, we get that $A^{(t)}_{0,0}$ projects onto the vector $w_t\in \bR^{k+1}$ defined by (where we assumed $\ell=k_1$):
\[
w_t\elem[\ell] = \sqrt{{n_1\choose \ell}{n_0\choose k-\ell}} \sum_{i=0}^t (-1)^{i} {\ell \choose i}{k-\ell \choose t-i} (n_1-\ell)^{\underline{t-i}}\;(n_0-k+\ell)^{\underline i}.
\]

\mycommand{tw}{\widetilde{w}}
Let $\tw_t = w_t/\|w_t\|$.
Assuming $n\to\infty$ and $n_1/n \to p$, it is easy to check that $\tw_t\to \varkappa_t^{(k,p)}$.
Also, the convergence is uniform if $c<p<1-c$.

Notice that the largest power of $n_0$ in $w_t\elem[\ell]$ is $(k-\ell)/2+\min\{\ell,t\}$, and the largest power of $n_1$ is $\ell/2 + \min\{k-\ell,t\}$.  Hence, $\tw_t$ is close to $e_t$ if $n_1/n$ is sufficiently small.  Also, $\tw_t$ is close to $e_{k-t}$ if $n_0/n$ is small.
\pfend

\section{Exact-Half Function}
\label{sec:exactHalf}
In this section, we apply \rf(thm:main) to the exact-half function.  The function $\text{EXACT-HALF}_k\colon \{0,1\}^k\to\{0,1\}$ is defined by $\text{EXACT-HALF}_k(x)=1$ iff $|x|=\lfloor k/2\rfloor$.  
This section is devoted to the proof of the following result:
\begin{thm}
\label{thm:exactHalf}
The quantum query complexity of $L_{\text{EXACT-HALF}_k}$ is $\Theta(k^{1/4})$.
\end{thm}

The lower bound can be shown using a simple positive-weighted adversary.  Consider the adversary matrix $\Gamma$ for $L_{\text{EXACT-HALF}_k}^n$ defined by $\Gamma\elem[A,B]=1$ if $A\ne B$ and $\Gamma\elem[A,A]=0$.  We have $\|\Gamma\| = {n\choose k}-1$.  
On the other hand, $\Gamma\circ\Delta_S$ is the all-1 matrix, and a simple argument involving~\rf(lem:binomial) shows that it has $O\sA[{n\choose k}/\sqrt{k}]$ columns.  
Also, $\Gamma\circ\Delta_S$ still has almost ${n\choose k}$ rows, hence, $\|\Gamma\circ\Delta_S\| = O\sA[{n\choose k}k^{-1/4}]$.  Thus, $\pAdv(L_h^n) = \Omega(k^{1/4})$.  In the remaining part of this section, we show that this simple lower bound is actually tight.

We do so by providing a feasible solution to the optimisation problem in \rf(thm:main).
Note that this optimisation problem has the following self-reducibility property:  For every $k'<k$, if we denote $d'_i = d_{i+k-k'}$ and take an appropriate subset of the constraints, we obtain the optimisation problem for the case $h=\text{EXACT-HALF}_{k'}$.
This has a number of consequences.  The first one is that it suffices to estimate $d_0$ only, because $d_i$ with larger values of $i$ have been already estimated for smaller values of $k$.

We only consider the constraints $\norm|M_{m, 1/2, \floor[k/2]-\floor[m/2]}^{(d)}|\le 1$, where $m$ ranges from $1$ to $k$.  
Let,
\begin{equation}
\label{eqn:AmlDef}
A_{m,\ell}=
\sA[\varkappa^{(m,1/2)}_\ell(\varkappa^{(m,1/2)}_\ell)^*]\elem[{W_0(t),W_1(t)}]
\end{equation}
in the notation of~\rf(eqn:Mdef) and~\rf(eqn:Mdef2), where $t=\floor[k/2]-\floor[m/2]$.
With this choice of parameters, $A_{m,\ell}$ is an $m\times 1$ matrix.
Later in the proof, we shall treat it as a vector in $\bR^m$.
Note also that the matrices $A_{m,\ell}$ do {\em not} depend on $k$.
Thus, we get the following optimization problem:
\begin{subequations}
\label{eqn:weightOriginalProblem}
\begin{alignat}{3}
 &{\mbox{\rm maximise }} &\qquad& d_0 \label{eqn:weightOriginalObjective}\\ 
 &{\mbox{\rm subject to }} && \normB|\sum_{i=1}^{m} d_{k-i} A_{m,m-i}|\le 1 &\qquad& \mbox{for all $m=1,\dots,k$;} \\
 &&& d_i\in \bR && \text{for $i=0,\dots,k-1$.}
\end{alignat}
\end{subequations}
Applying semi-definite duality~\cite[Section 5.9]{boyd:convex}, we obtain the following upper bound on~\rf(eqn:weightOriginalProblem):
\begin{subequations}
\label{eqn:weightOptimize}
\begin{alignat}{3}
 &{\mbox{\rm minimise }} &\qquad& \sum_{m=1}^{k} \normtr|\Lambda_m| \label{eqn:objective}\\ 
 &{\mbox{\rm subject to }} && \ip<\Lambda_k, A_{k,0}>=1; \label{eqn:first} \\
 &&& \sum_{i=0}^\ell \ip<\Lambda_{k-i}, A_{k-i, \ell-i}>=0 &\qquad& \mbox{for all $\ell=1,\dots,k-1$}; \label{eqn:second}\\
 &&& \text{$\Lambda_{m}$ has the same size as $A_{m,\ell}$} && \text{for $m=1,\dots,k$ and any $\ell$.}
\end{alignat}
\end{subequations}

We are going to construct a feasible solution to this problem.  We use the following elimination strategy.  Constraint~\rf(eqn:first) only uses $\Lambda_k$.  So, we take some $\Lambda_k$ that satisfies $\ip<\Lambda_k, A_{k,0}>=1$, but may have non-zero inner products with other $A_{k,i}$.  Then we take $\Lambda_{k-1}$ that satisfies~\rf(eqn:second) for $\ell=1$, then $\Lambda_{k-2}$ that satisfies~\rf(eqn:second) for $\ell=2$, and so on.  We find $\Lambda_{k-i}$ for $i>0$ using self-reducibility.

More formally, we apply induction.  Let $\Lambda^{(k)} = (\Lambda^{(k)}_1,\dots, \Lambda^{(k)}_k)$ be our solution to~\rf(eqn:weightOptimize) for a specific value of $k$, 
and let $g(k)$ denote the corresponding value of~\rf(eqn:objective).  The base case, $\Lambda^{(1)}$, is trivial to construct.  Assume we have constructed $\Lambda^{(k')}$ for all $k'<k$.  Then, we take 
\[
\Lambda^{(k)} = (0,\dots,0,\Lambda_k) - \sum_{\ell=1}^{k-1} \ip<\Lambda_k, A_{k,\ell}> \Lambda^{(k-\ell)},
\]
where the first list has $\Lambda_k$ in the $k$th position, and the remaining lists are padded with zeroes from the right.  
Here $\Lambda_k$ is some matrix satisfying~\rf(eqn:first).  We shall define it later.  It is easy to check that $\Lambda^{(k)}$ satisfies~\rf(eqn:first) and~\rf(eqn:second).  Using the triangle inequality for the trace norm, we obtain
\begin{equation}
\label{eqn:gRekursija}
g(k) \le \normtr|\Lambda_k| + \sum_{\ell=1}^{k-1} \absA|\ip<\Lambda_k, A_{k,\ell}>|\; g(k-\ell).
\end{equation}

\mycommand{dyrkappa}{\breve\varkappa}
So, it remains to choose $\Lambda_k$.  
For the remainder of this section and the next section, let $\varkappa_\ell = \varkappa^{(k,1/2)}_\ell$.  Recall that $\{\varkappa_\ell\}$ form an orthonormal basis of $\bR^{k+1}$.
Let, for brevity, $s = \floor[k/2]$.  We have $A_{k,\ell} = \varkappa_\ell\elem[s]\; \dyrkappa_\ell$, where $\dyrkappa_\ell$ denotes $\varkappa_\ell$ with the $s$th element removed.  We take
\[
\Lambda_k = \frac{1}{\varkappa_0\elem[s](1-\varkappa_0\elem[s]^2)}\;\dyrkappa_0.
\]
It is straightforward to check that $\ip<\Lambda_k, A_{k,0}>=1$.  
Also, for $\ell>0$,
\(
\ip<\dyrkappa_0, A_{k,\ell}> = \varkappa_\ell\elem[s]\ip<\dyrkappa_0, \dyrkappa_\ell> = - \varkappa_0\elem[s]\varkappa_\ell\elem[s]^2.
\)
Hence,
\begin{equation}
\label{eqn:LambdaAkell}
\ip<\Lambda_k, A_{k,\ell}> = \frac{-\varkappa_\ell\elem[s]^2}{1-\varkappa_0\elem[s]^2}.
\end{equation}
Now we apply some additional properties of $\varkappa_\ell$.  First of all,
\(
\varkappa_0\elem[x] = \sqrt{{k\choose x}/2^k}.
\)
Thus, by \rf(lem:binomial), $\varkappa_0\elem[s] = \Theta(k^{-1/4})$, and $\normtr|\Lambda_k| = \Theta(k^{1/4})$.

Another property~\cite[Eq.~(32)]{krasikov:KrawtchoukSurvey} is $\varkappa_\ell\elem[s] = \pm \varkappa_{k-\ell}\elem[s]$.  As $\{\varkappa_\ell\}$ form an orthonormal basis, we get $\sum_{\ell=0}^k \varkappa_\ell\elem[s]^2=1$, hence, by~\rf(eqn:LambdaAkell),
\[
\sum_{\ell=1}^{k-1} |\ip<\Lambda_k, A_{k,\ell}>| = \frac{1-2\varkappa_0\elem[s]^2}{1-\varkappa_0\elem[s]^2},
\qquad\mbox{and}\qquad
\frac{1-\varkappa_0\elem[s]^2}{1-2\varkappa_0\elem[s]^2} \sum_{\ell=1}^{k-1} (k-\ell)|\ip<\Lambda_k, A_{k,\ell}>| = \frac k2.
\]
Let $C_0$ be some constant such that $g(k)\le C_0 k^{1/4}$ for small values of $k$, and let $C_1$ be such that $\normtr|\Lambda_k|\le C_1 k^{1/4}$ for all $k$.  Then, we prove by induction that $g(k)\le Ck^{1/4}$ for $C=\max\{C_0, \frac{\sqrt[4]{2}}{\sqrt[4]{2}-1} C_1\}$.  Indeed, this is satisfied for the small values of $k$.  Assume this is satisfied for all $k'<k$.  Then, by~\rf(eqn:gRekursija) and the concavity of $k^{1/4}$:
\begin{align*}
g(k) \le C_1k^{1/4} + \sum_{\ell=1}^{k-1} |\ip<\Lambda_k, A_{k,\ell}>| C(k-\ell)^{1/4} 
&\le C_1k^{1/4} + C \sC[\frac{1-\varkappa_0\elem[s]^2}{1-2\varkappa_0\elem[s]^2} \sum_{\ell=1}^{k-1} |\ip<\Lambda_k, A_{k,\ell}>| (k-\ell)]^{1/4}\\
&= C_1k^{1/4} + C(k/2)^{1/4} \le C k^{1/4}.
\end{align*}

\section{Majority Function}
\label{sec:majority}
In this section, we prove some partial results on the quantum query complexity of the $L_{\text{MAJORITY}_k}$ function.  The function is defined by $\text{MAJORITY}_k(x)=1$ iff $|x|\ge k/2$.  First, the algorithm from \rf(sec:exactHalf) carries over to this case with minor modifications.

\begin{thm}
\label{thm:majority}
The quantum query complexity of $L_{\text{MAJORITY}_k}$ is $O(k^{1/4})$.
\end{thm}

\pfstart
Again, we construct a feasible solution to~\rf(eqn:weightOptimize) where $A_{m,\ell}$ are as in~\rf(eqn:AmlDef) with $W_0$ and $W_1$ modified accordingly.  This time, we use different strategies to construct $\Lambda^{(k)}$ for odd and even values of $k$.  We need the following easy symmetry result about Krawtchouk polynomials~\cite[Eq.~(31)]{krasikov:KrawtchoukSurvey}:
\begin{equation}
\label{eqn:krawtchoukSymmetry}
\varkappa_\ell\elem[x] = (-1)^\ell\varkappa_\ell\elem[k-x],
\end{equation}
where again $\varkappa_\ell = \varkappa_\ell^{(k,1/2)}$.  We also use notations $W_0 = W_0(0)$ and $W_1 = W_1(0)$.

For the even values of $k$, we use the same elimination strategy, but we change the way we define the matrix $\Lambda_k$.  Let $s=k/2$, and let this time $\dyrkappa_\ell$ denote the $W_0\times W_1$ matrix having the elements of $\varkappa\elem[W_0]$ in column $s$, and zeroes everywhere else.  Intuitively, the non-zero elements of $\dyrkappa_\ell$ form the upper half of the vector $\dyrkappa_\ell$ from the proof of \rf(thm:exactHalf).  We define
\[
\Lambda_k = \frac{2}{\varkappa_0\elem[s](1-\varkappa_0\elem[s]^2)}\; \dyrkappa_0.
\]
From~\rf(eqn:krawtchoukSymmetry), we get $\ip<\Lambda_k, A_{k,0}>=1$.  Also, $\normtr|\Lambda_k| = \Theta(k^{1/4})$.
If $\ell$ is odd we get from~\rf(eqn:krawtchoukSymmetry) that the $s$th column of $A_{k,\ell}$ consists of zeroes.  If $\ell$ is even, using the same property, we get that $\ip<\dyrkappa_0,\dyrkappa_\ell> = -\varkappa_0\elem[s]\varkappa_\ell\elem[s]/2$.  Either way,~\rf(eqn:LambdaAkell) holds.  The proof further proceeds as in \rf(sec:exactHalf).  Also, we have to note that $\ip<\Lambda_k, A_{k,\ell}>=0$ if $\ell$ is odd, hence, we only need $\Lambda^{(k')}$ with even values of $k'<k$ to define $\Lambda^{(k)}$.

Now assume that $k$ is odd.  We know that $d_1 = O(k^{1/4})$ by considering $\Lambda^{(k-1)}$.  
Thus, we change our strategy and prove that $d_0-d_1=O(1)$.
If we replace~\rf(eqn:weightOriginalObjective) by $d_0-d_1$, we get the problem~\rf(eqn:weightOptimize) with~\rf(eqn:first) replaced by
\begin{equation}
\label{eqn:third}
\ip<\Lambda_k, A_{k,0}>=1\qquad\mbox{and}\qquad \ip<\Lambda_k, A_{k,1}> + \ip<\Lambda_{k-1}, A_{k-1,0}> = -1,
\end{equation}
and $\ell$ ranging in~\rf(eqn:second) from 2 to $k-1$.  A possible feasible solution is
\[
\Lambda_k = \frac{2}{\ip<\varkappa_0\elem[W_1],\varkappa_1\elem[W_1]>}(\varkappa_0\elem[W_0])(\varkappa_1\elem[W_1])^*
\]
and $\Lambda_{m}=0$ for other values of $m$.  Using~\rf(eqn:krawtchoukSymmetry) and the orthogonality of $\{\varkappa_\ell\}$, we get that $\ip<\varkappa_0\elem[W_0],\varkappa_\ell[W_0]>=0$ for even $\ell\ge 2$, and $\ip<\varkappa_1\elem[W_1],\varkappa_\ell[W_1]>=0$ for odd $\ell\ge 3$, hence~\rf(eqn:second) holds.  We get~\rf(eqn:third) similarly.  Finally, $K_1\elem[x] = k-2x$, $\|K_0\|=1$ and $\|K_1\|=\sqrt{k}$, where $K$ is defined in~\rf(eqn:Kdef) \cite[Eqs.~(12, 33)]{krasikov:KrawtchoukSurvey}, thus, using the definition of $\varkappa$ and the central limit theorem:
\[
\ip<\varkappa_0\elem[W_1],\varkappa_1\elem[W_1]> = \frac{1}{2^{k}\sqrt{k}} \sum_{x=\ceil[k/2]}^k {k\choose x}(k-2x) \stackrel{k\to\infty}{\longrightarrow} -\frac{4}{k\sqrt{2\pi}} \int_{0}^\infty \ee^{-2x^2/k}x\;\dd x = -\frac{1}{\sqrt{2\pi}}.
\]
Hence, $\normtr|\Lambda_k|=O(1)$.
\pfend

For the case when $h$ is the OR or the exact-half function, we were able to prove tight lower bounds using the positive-weighted adversary.  In the next theorem, we show that it is not possible to prove a polynomial (in $k$) lower bound using this technique, when $h$ is the majority function.  There are some limitations known on the positive-weighted adversary, like the certificate complexity barrier~\cite{spalek:advEquivalent} or the property testing barrier~\cite{hoyer:advNegative}.  Neither apply here, so we give a direct proof using the optimisation problem given by~\rf(eqn:advDualObjective),~\rf(eqn:advPositive) and~\rf(eqn:advDualSemidefinite).

\begin{thm}
The positive-weighted adversary bound $\pAdv(L_{\text{MAJORITY}_k})$ is $O(\log k)$.
\end{thm}

\pfstart
Fix $n$.  If $X=(X_S)$ is a family of positive semi-definite matrices, let $m(X)$ stand for the objective~\rf(eqn:advDualObjective), and $\ell_{A,B}(X)$ stand for the left-hand side of~\rf(eqn:advPositive).  The proof is based on the following lemma:
\begin{lem}
\label{lem:advMaj1}
For each $1\le d\le k$, there exist positive semi-definite matrices $X=(X_S)$ with non-negative entries such that $m(X)=O(1)$ and $\ell_{A,B}(X) \ge 1$ for all $A, B\in\cC$ satisfying $d \le |A\setminus B|\le 2d$.
\end{lem}

The theorem immediately follows from \rf(lem:advMaj1).  Indeed, we cover the interval $[1,k]$ with a logarithmic number of intervals of the form $[d,2d]$.  For each of them, we apply \rf(lem:advMaj1) and take the sum of the resulting matrices.

So, it remains to prove the lemma.  Consider the matrices $X_S$ built in the following way.  For each $S$, $X_S$ is a rank-1 matrix with $X_S\elem[A,B]=2^{-n}$ if both $|A\cap S|$ and $|B\cap S|$ lie in the interval $[k/2-\sqrt{d},\; k/2+\sqrt{d}]$, and zeroes elsewhere.  Using \rf(lem:binomial), we get that, for all $A$,
\begin{equation}
\label{eqn:majAdv1}
m(X) = \pr_S\skB[\frac k2-\sqrt{d}\le |S\cap A|\le \frac k2+\sqrt{d}] = \Theta(\sqrt{d/k}),
\end{equation}
where $S$ is taken uniformly at random from $2^{[n]}$.  Fix $A,B\in\cC$, and let $\ell=|A\setminus B|$.  Assume $d\le \ell\le 2d$.  Again, we have $\Pr_S\skA[ \frac{k-\ell}2 - \sqrt{d} \le |A\cap B \cap S|\le \frac{k-\ell}2 + \sqrt{d} ] = \Omega(\sqrt{d/k})$.  Also, provided that the last condition on $A\cap B\cap S$ holds, we get that $\frac k2-\sqrt{d} \le |A\cap S|<\frac k2$ with probability $\Omega(1)$, and similarly for $\frac k2 \le |B\cap S| \le \frac k2+\sqrt{d}$.  Thus,
\[
\ell_{A,B}(X) = \Omega(\sqrt{d/k}).
\]
Combining this with~\rf(eqn:majAdv1), and rescaling the matrices $X_S$, we get the statement of \rf(lem:advMaj1).
\pfend

\section{Further Observations}
\label{sec:nonadaptive}

In this section, we prove two additional results about the problems studied in the previous sections.
First, we show that many of the above algorithms can be made exact.

\begin{prp}
The quantum algorithms for $L_{\text{OR}_k}$, $L_{\text{EXACT-HALF}_k}$ and $L_{\text{MAJORITY}_k}$ from Theorems~\ref{thm:combGroupTest}, \ref{thm:exactHalf} and~\ref{thm:majority} can be made exact without increasing their complexity.
\end{prp}

\pfstart
We use the same observation as in~\cite{iwama:quantumCounterfeit}.  
Inputs to all these problems are $k$-subsets of $[n]$.  Due to symmetry, the error probability of any of these algorithms is the same on all inputs.  Also, for each of the problems, there exists a deterministic procedure that efficiently tests whether a given $k$-subset $A$ is the true input.  Indeed, for the OR function, query the complement of $A$.  For exact-half, cover $A$ by 3 subsets of size $\floor[k/2]$ and query each of them.  Similarly for the majority function.

This means that we can apply the exact amplitude amplification algorithm from~\cite{brassard:amplification}, and get an exact algorithm with an $O(1)$ multiplicative overhead in complexity.
\pfend

Next, we show that the query complexity achieved in the previous sections cannot be obtained by a non-adaptive quantum query algorithm.  If $h$ is the OR function, any non-adaptive quantum algorithm requires $\Omega(k)$ queries.  This follows from Zalka's result~\cite{zalka:GroverOptimal} and the fact that unstructured search can be reduced to $L_{\text{OR}}$.  For the remaining problems, we obtain the following result:

\begin{thm}
\label{thm:nonadaptive}
The non-adaptive quantum query complexity of $L_{\text{\rm MAJORITY}_k}$ and $L_{\text{\rm EXACT-HALF}_k}$ is $\Omega(\sqrt{k})$.
\end{thm}

Note that this result is nearly tight:  Using Fourier sampling like in~\cite{atici:testingJuntas}, it is possible to solve both problems in $\tilde O(\sqrt{k})$ quantum queries non-adaptively.
Indeed, the Fourier spectrum of the majority and the exact-half functions is concentrated on sets of size roughly $\sqrt{k}$, so, after $O(\sqrt{k}\log k)$ Fourier samples, it is likely to have seen all $k$ relevant variables.

\pfstart[Proof of \rf(thm:nonadaptive)]
Essentially, we use the non-adaptive version of the adversary bound from~\cite{koiran:nonAdaptiveAdversary}.  We give a direct proof, however.
Consider a non-adaptive $T$-query algorithm for one of these problems on $n$ variables.  The state of the algorithm before the query is of the form
\[
\psi = \sum_{S_1,\dots,S_T} \alpha_{S_1,\dots,S_T}\ket|S_1,\dots,S_T>\ket|\phi_{S_1,\dots,S_T}>,
\]
where $S_i$ are subsets of $[n]$, and $\phi_{S_1,\dots,S_T}$ are some unit vectors.
Assuming $T=o(\sqrt{k})$, we are going to construct two subsets $A$ and $B$ such that $O_A^{\otimes T}\psi$ and $O_B^{\otimes T}\psi$ have large inner product.  For the latter, we have
\begin{equation}
\label{eqn:OAOBinner}
\absA|\ipA<O_A^{\otimes T}\psi, O_B^{\otimes T}\psi>| \ge 2 \sum |\alpha_{S_1,\dots,S_T}|^2 - 1,
\end{equation}
where the summation is over all $(S_1,\dots,S_T)$ such that $f_A(S_i)=f_B(S_i)$ for all $i\in[T]$.

The subsets $A$ and $B$ will be such that $A\cap B = D$ with $|D|=k-1$.  Then, $f_A(S)= f_B(S)$ if $|S\cap D|\notin \{\lceil k/2 \rceil-1,  \lceil k/2 \rceil \}$ for both cases of $h$ equal to MAJORITY$_k$ or EXACT-HALF$_k$.

It is easy to show, using \rf(lem:binomial), that if $D$ is a $(k-1)$-subset of $[n]$ taken uniformly at random, and $n$ is large enough, then, for any $S\subseteq[n]$, the probability of $|S\cap D| \in \{\lceil k/2 \rceil-1,  \lceil k/2 \rceil \}$ is  $O(1/\sqrt{k})$.  By the union bound, the probability that $f_A(S_i)=f_B(S_i)$ for all $i\in[T]$ is $1-o(1)$.  By the linearity of expectation, the expectation of the right-hand side of~\rf(eqn:OAOBinner) is $1-o(1)$.  Hence, there exist $A$ and $B$ such that it is not possible to distinguish $O_A^{\otimes T}\psi$ and $O_B^{\otimes T}\psi$ with error probability less than $1/3$.
\pfend

\section{Discussion}
In this paper, we studied the quantum query complexity of the function $L_h$, when $h$ is the OR, the exact-half, and the majority function.  For the first two functions, we gave optimal algorithms.  The algorithms are based on the adversary bound, and attain at least quartic improvement in query complexity in comparison to the randomised algorithms when $h$ is the exact-half or the majority function.  This shows that the dual adversary bound can be an important tool for quantum learning algorithms.

One apparent open problem is the study of $Q(L_h)$ for other functions $h$.
For instance, can our solution in \rf(sec:exactHalf) be generalised to the exact-$\ell$ or the $\ell$-threshold functions?  
For the majority function, there is still an exponential gap between the lower and the upper bounds that we can prove.  If the query complexity is logarithmic, we would get an exponential separation using quantum walks.  There is already an example of such separation~\cite{childs:walkExponentialSeparation}, but the problem studied in the latter paper is not so natural.  However, we believe that the complexity is polynomial in $k$.  In this case, we would get an example of a quantum query lower bound outperforming the positive-weighted adversary.  There are not so many cases known when a general adversary is strictly better than a positive-weighted adversary~\cite{reichardt:formulae, belovs:onThePower}.  Of course, it is also possible to use the polynomial method, as was done for the collision problem~\cite{shi:collisionLower}.

Another open problem is to use these ideas in the development of other learning or property testing algorithms.  For instance, the combinatorial group testing problem is related to junta testing.  Is it possible to use any ideas from the current paper to improve the algorithm in~\cite{atici:testingJuntas}?

\subsection*{Acknowledgements}
I am grateful to Ashley Montanaro for introducing me to this problem and for helpful advice.  I would also like to thank Andris Ambainis, Mihails Belovs, Ansis Rosmanis for useful discussions, and Oded Regev for pointing out the reference~\cite{iwama:quantumCounterfeit}.
I thank anonymous referees for many useful suggestions on improving presentation of the paper.

The author is supported by Scott Aaronson's Alan T. Waterman Award from the National Science Foundation.  This research was performed when the author was at the University of Latvia and was supported by the European Social Fund within the project ``Support for Doctoral Studies at University of Latvia'' and by ERC Advanced Grant MQC.

\bibliographystyle{../../habbrvM}
\bibliography{../../bib}

\begin{thebibliography}{10}

\bibitem{shi:collisionLower}
S.~Aaronson and Y.~Shi.
\newblock Quantum lower bounds for the collision and the element distinctness
  problems.
\newblock {\em Journal of the ACM},
  51(4):\myhref{http://dx.doi.org/10.1145/1008731.1008735}{595--605}, 2004.

\bibitem{ambainis:adv}
A.~Ambainis.
\newblock Quantum lower bounds by quantum arguments.
\newblock {\em Journal of Computer and System Sciences},
  64(4):\myhref{http://dx.doi.org/10.1006/jcss.2002.1826}{750--767}, 2002.
\newblock Earlier: \myhref{http://dx.doi.org/10.1145/335305.335394}{{\em
  STOC'00}},  \myhref{http://arxiv.org/abs/quant-ph/0002066}{{\ttfamily
  arXiv:quant-ph/0002066}}.

\bibitem{ambainis:oracleIdentification}
A.~Ambainis, K.~Iwama, A.~Kawachi, H.~Masuda, R.~H. Putra, and S.~Yamashita.
\newblock Quantum identification of boolean oracles.
\newblock In {\em Proc. of 21st STACS}, volume 2996 of {\em LNCS}, pages
  \myhref{http://dx.doi.org/10.1007/978-3-540-24749-4_10}{105--116}. Springer,
  2004.
\newblock  \myhref{http://arxiv.org/abs/quant-ph/0403056}{{\ttfamily
  arXiv:quant-ph/0403056}}.

\bibitem{montanaro:groupTesting}
A.~Ambainis and A.~Montanaro.
\newblock Quantum algorithms for search with wildcards and combinatorial group
  testing.
\newblock {\em Quantum Information \& Computation}, 14(5\&6):439--453, 2014.
\newblock  \myhref{http://arxiv.org/abs/1210.1148}{{\ttfamily
  arXiv:1210.1148}}.

\bibitem{angluin:conceptLearning}
D.~Angluin.
\newblock Queries and concept learning.
\newblock {\em Machine learning},
  2(4):\myhref{http://dx.doi.org/10.1007/BF00116828}{319--342}, 1988.

\bibitem{atici:improvedBoundsLearning}
A.~At{\i}c{\i} and R.~A. Servedio.
\newblock Improved bounds on quantum learning algorithms.
\newblock {\em Quantum Information Processing},
  4(5):\myhref{http://dx.doi.org/10.1007/s11128-005-0001-2}{355--386}, 2005.
\newblock  \myhref{http://arxiv.org/abs/quant-ph/0411140}{{\ttfamily
  arXiv:quant-ph/0411140}}.

\bibitem{atici:testingJuntas}
A.~At{\i}c{\i} and R.~A. Servedio.
\newblock Quantum algorithms for learning and testing juntas.
\newblock {\em Quantum Information Processing},
  6(5):\myhref{http://dx.doi.org/10.1007/s11128-007-0061-6}{323--348}, 2007.
\newblock  \myhref{http://arxiv.org/abs/0707.3479}{{\ttfamily
  arXiv:0707.3479}}.

\bibitem{belovs:learningKDist}
A.~Belovs.
\newblock Learning-graph-based quantum algorithm for $k$-distinctness.
\newblock In {\em Proc. of 53rd IEEE FOCS}, pages
  \myhref{http://dx.doi.org/10.1109/FOCS.2012.18}{207--216}, 2012.
\newblock  \myhref{http://arxiv.org/abs/1205.1534}{{\ttfamily
  arXiv:1205.1534}}.

\bibitem{belovs:learning}
A.~Belovs.
\newblock Span programs for functions with constant-sized 1-certificates.
\newblock In {\em Proc. of 44th ACM STOC}, pages
  \myhref{http://dx.doi.org/10.1145/2213977.2213985}{77--84}, 2012.
\newblock  \myhref{http://arxiv.org/abs/1105.4024}{{\ttfamily
  arXiv:1105.4024}}.

\bibitem{belovs:learningClaws}
A.~Belovs and B.~W. Reichardt.
\newblock Span programs and quantum algorithms for $st$-connectivity and claw
  detection.
\newblock In {\em Proc. of 20th ESA}, volume 7501 of {\em LNCS}, pages
  \myhref{http://dx.doi.org/10.1007/978-3-642-33090-2_18}{193--204}, 2012.
\newblock  \myhref{http://arxiv.org/abs/1203.2603}{{\ttfamily
  arXiv:1203.2603}}.

\bibitem{belovs:onThePower}
A.~Belovs and A.~Rosmanis.
\newblock On the power of non-adaptive learning graphs.
\newblock {\em Computational Complexity},
  23(2):\myhref{http://dx.doi.org/10.1007/s00037-014-0084-1}{323--354}, 2014.
\newblock Earlier: \myhref{http://dx.doi.org/10.1109/CCC.2013.14}{{\em
  CCC'13}},  \myhref{http://arxiv.org/abs/1210.3279}{{\ttfamily
  arXiv:1210.3279}}.

\bibitem{bernstein:quantumComplexity}
E.~Bernstein and U.~Vazirani.
\newblock Quantum complexity theory.
\newblock {\em SIAM Journal on Computing},
  26(5):\myhref{http://dx.doi.org/10.1137/S0097539796300921}{1411--1473}, 1997.
\newblock Earlier: \myhref{http://dx.doi.org/10.1145/167088.167097}{{\em
  STOC'93}}.

\bibitem{blais:testingJuntas}
E.~Blais.
\newblock Testing juntas nearly optimally.
\newblock In {\em Proc. of 41st ACM STOC}, pages
  \myhref{http://dx.doi.org/10.1145/1536414.1536437}{151--158}, 2009.

\bibitem{boyd:convex}
S.~Boyd and L.~Vandenberghe.
\newblock {\em Convex optimization}.
\newblock Cambridge University Press, 2004.

\bibitem{brassard:amplification}
G.~Brassard, P.~H{\o}yer, M.~Mosca, and A.~Tapp.
\newblock Quantum amplitude amplification and estimation.
\newblock In {\em Quantum Computation and Quantum Information: A Millennium
  Volume}, volume 305 of {\em AMS Contemporary Mathematics Series}, pages
  53--74, 2002.
\newblock  \myhref{http://arxiv.org/abs/quant-ph/0005055}{{\ttfamily
  arXiv:quant-ph/0005055}}.

\bibitem{bshouty:oraclesForExactLearning}
N.~H. Bshouty, R.~Cleve, R.~Gavald{\`a}, S.~Kannan, and C.~Tamon.
\newblock Oracles and queries that are sufficient for exact learning.
\newblock {\em Journal of Computer and System Sciences},
  52(3):\myhref{http://dx.doi.org/10.1006/jcss.1996.0032}{421--433}, 1996.
\newblock Earlier: \myhref{http://dx.doi.org/10.1145/180139.181067}{{\em
  COLT'94}}.

\bibitem{bshouty:learningDNF}
N.~H. Bshouty and J.~C. Jackson.
\newblock Learning {DNF} over the uniform distribution using a quantum example
  oracle.
\newblock {\em SIAM Journal on Computing},
  28(3):\myhref{http://dx.doi.org/10.1137/S0097539795293123}{1136--1153}, 1998.
\newblock Earlier: \myhref{http://dx.doi.org/10.1145/225298.225312}{{\em
  COLT'95}}.

\bibitem{buhrman:querySurvey}
H.~Buhrman and R.~de~Wolf.
\newblock Complexity measures and decision tree complexity: a survey.
\newblock {\em Theoretical Computer Science},
  288:\myhref{http://dx.doi.org/10.1016/S0304-3975(01)00144-X}{21--43}, 2002.

\bibitem{childs:walkExponentialSeparation}
A.~M. Childs, R.~Cleve, E.~Deotto, E.~Farhi, S.~Gutmann, and D.~A. Spielman.
\newblock Exponential algorithmic speedup by a quantum walk.
\newblock In {\em Proc. of 35th ACM STOC}, pages
  \myhref{http://dx.doi.org/10.1145/780542.780552}{59--68}, 2003.
\newblock  \myhref{http://arxiv.org/abs/quant-ph/0209131}{{\ttfamily
  arXiv:quant-ph/0209131}}.

\bibitem{childs:hiddenShift}
A.~M. Childs, R.~Kothari, M.~Ozols, and M.~R{\"o}tteler.
\newblock Easy and hard functions for the boolean hidden shift problem.
\newblock In {\em Proc. of 8th TQC}, volume~22 of {\em LIPIcs}, pages
  \myhref{http://dx.doi.org/10.4230/LIPIcs.TQC.2013.50}{50--79}. Dagstuhl,
  2013.
\newblock  \myhref{http://arxiv.org/abs/1304.4642}{{\ttfamily
  arXiv:1304.4642}}.

\bibitem{curtis:representationTheory}
C.~W. Curtis and I.~Reiner.
\newblock {\em Representation theory of finite groups and associative
  algebras}.
\newblock AMS, 1962.

\bibitem{du:combinatorialGroupTesting}
D.~Z. Du and F.~Hwang.
\newblock {\em Combinatorial group testing and its applications}, volume~3 of
  {\em Series on Applied Mathematics}.
\newblock World Scientific, 1993.

\bibitem{ettinger:hspQuery}
M.~Ettinger, P.~H{\o}yer, and E.~Knill.
\newblock The quantum query complexity of the hidden subgroup problem is
  polynomial.
\newblock {\em Information Processing Letters},
  91(1):\myhref{http://dx.doi.org/10.1016/j.ipl.2004.01.024}{43--48}, 2004.
\newblock  \myhref{http://arxiv.org/abs/quant-ph/0401083}{{\ttfamily
  arXiv:quant-ph/0401083}}.

\bibitem{hausladen:PGM}
P.~Hausladen and W.~K. Wootters.
\newblock A 'pretty good' measurement for distinguishing quantum states.
\newblock {\em Journal of Modern Optics},
  41(12):\myhref{http://dx.doi.org/10.1080/09500349414552221}{2385--2390},
  1994.

\bibitem{hoyer:advNegative}
P.~H{\o}yer, T.~Lee, and R.~{\v Spalek}.
\newblock Negative weights make adversaries stronger.
\newblock In {\em Proc. of 39th ACM STOC}, pages
  \myhref{http://dx.doi.org/10.1145/1250790.1250867}{526--535}, 2007.
\newblock  \myhref{http://arxiv.org/abs/quant-ph/0611054}{{\ttfamily
  arXiv:quant-ph/0611054}}.

\bibitem{iwama:quantumCounterfeit}
K.~Iwama, H.~Nishimura, R.~Raymond, and J.~Teruyama.
\newblock Quantum counterfeit coin problems.
\newblock {\em Theoretical Computer Science},
  456:\myhref{http://dx.doi.org/10.1016/j.tcs.2012.05.039}{51--64}, 2012.
\newblock Earlier: \myhref{http://dx.doi.org/10.1007/978-3-642-17517-6_10}{{\em
  ISAAC'10}},  \myhref{http://arxiv.org/abs/1009.0416}{{\ttfamily
  arXiv:1009.0416}}.

\bibitem{koiran:nonAdaptiveAdversary}
P.~Koiran, J.~Landes, N.~Portier, and P.~Yao.
\newblock Adversary lower bounds for nonadaptive quantum algorithms.
\newblock {\em Journal of Computer and System Sciences},
  76(5):\myhref{http://dx.doi.org/10.1016/j.jcss.2009.10.007}{347--355}, 2010.
\newblock Earlier: \myhref{http://dx.doi.org/10.1007/978-3-540-69937-8_20}{{\em
  WoLLIC'08}},  \myhref{http://arxiv.org/abs/0804.1440}{{\ttfamily
  arXiv:0804.1440}}.

\bibitem{kothari:oracleIdentification}
R.~Kothari.
\newblock An optimal quantum algorithm for the oracle identification problem.
\newblock In {\em Proc. of 31st STACS}, volume~25 of {\em LIPIcs}, pages
  \myhref{http://dx.doi.org/10.4230/LIPIcs.STACS.2014.482}{482--493}. Dagstuhl,
  2014.
\newblock  \myhref{http://arxiv.org/abs/1311.7685}{{\ttfamily
  arXiv:1311.7685}}.

\bibitem{krasikov:KrawtchoukSurvey}
I.~Krasikov and S.~Litsyn.
\newblock Survey of binary {K}rawtchouk polynomials.
\newblock In {\em Codes and association schemes}, volume~56 of {\em DIMACS
  series in Discrete Mathematics and Theoretical Computer Science}, pages
  199--212. AMS, 2001.

\bibitem{lee:learningTriangle}
T.~Lee, F.~Magniez, and M.~Santha.
\newblock Improved quantum query algorithms for triangle finding and
  associativity testing.
\newblock In {\em Proc. of 24th ACM-SIAM SODA}, pages
  \myhref{http://dx.doi.org/10.1137/1.9781611973105.107}{1486--1502}, 2013.
\newblock  \myhref{http://arxiv.org/abs/1210.1014}{{\ttfamily
  arXiv:1210.1014}}.

\bibitem{lee:stateConversion}
T.~Lee, R.~Mittal, B.~W. Reichardt, R.~{\v Spalek}, and M.~Szegedy.
\newblock Quantum query complexity of state conversion.
\newblock In {\em Proc. of 52nd IEEE FOCS}, pages
  \myhref{http://dx.doi.org/10.1109/FOCS.2011.75}{344--353}, 2011.
\newblock  \myhref{http://arxiv.org/abs/1011.3020}{{\ttfamily
  arXiv:1011.3020}}.

\bibitem{montanaro:nonadaptive}
A.~Montanaro.
\newblock Nonadaptive quantum query complexity.
\newblock {\em Information Processing Letters},
  110(24):\myhref{http://dx.doi.org/10.1016/j.ipl.2010.09.009}{1110--1113},
  2010.
\newblock  \myhref{http://arxiv.org/abs/1001.0018}{{\ttfamily
  arXiv:1001.0018}}.

\bibitem{reichardt:spanPrograms}
B.~W. Reichardt.
\newblock Span programs and quantum query complexity: The general adversary
  bound is nearly tight for every boolean function.
\newblock In {\em Proc. of 50th IEEE FOCS}, pages
  \myhref{http://dx.doi.org/10.1109/FOCS.2009.55}{544--551}, 2009.
\newblock  \myhref{http://arxiv.org/abs/0904.2759}{{\ttfamily
  arXiv:0904.2759}}.
\newblock Citations are to the arXiv version.

\bibitem{reichardt:formulae}
B.~W. Reichardt and R.~{\v Spalek}.
\newblock Span-program-based quantum algorithm for evaluating formulas.
\newblock {\em Theory of Computing},
  8:\myhref{http://dx.doi.org/10.4086/toc.2012.v008a013}{291--319}, 2012.
\newblock Earlier: \myhref{http://dx.doi.org/10.1145/1374376.1374394}{{\em
  STOC'08}},  \myhref{http://arxiv.org/abs/0710.2630}{{\ttfamily
  arXiv:0710.2630}}.

\bibitem{sagan:symmetricGroup}
B.~E. Sagan.
\newblock {\em The symmetric group: representations, combinatorial algorithms,
  and symmetric functions}, volume 203 of {\em Graduate Texts in Mathematics}.
\newblock Springer, 2001.

\bibitem{serre:representation}
J.-P. Serre.
\newblock {\em Linear Representations of Finite Groups}, volume~42 of {\em
  Graduate Texts in Mathematics}.
\newblock Springer, 1977.

\bibitem{servedio:equivalencesQuantum}
R.~A. Servedio and S.~J. Gortler.
\newblock Equivalences and separations between quantum and classical
  learnability.
\newblock {\em SIAM Journal on Computing},
  33(5):\myhref{http://dx.doi.org/10.1137/S0097539704412910}{1067--1092}, 2004.

\bibitem{spalek:advEquivalent}
R.~{\v Spalek} and M.~Szegedy.
\newblock All quantum adversary methods are equivalent.
\newblock {\em Theory of Computing},
  2:\myhref{http://dx.doi.org/10.4086/toc.2006.v002a001}{1--18}, 2006.
\newblock Earlier: \myhref{http://dx.doi.org/10.1007/11523468_105}{{\em
  ICALP'05}},  \myhref{http://arxiv.org/abs/quant-ph/0409116}{{\ttfamily
  arXiv:quant-ph/0409116}}.

\bibitem{szego:orthogonal}
G.~Szeg{\H{o}}.
\newblock {\em Orthogonal polynomials}, volume~23 of {\em AMS Colloquium
  Publications}.
\newblock 1975.

\bibitem{vanDam:oracleInterrogation}
W.~van Dam.
\newblock Quantum oracle interrogation: {G}etting all information for almost
  half the price.
\newblock In {\em Proc. of 39th IEEE FOCS}, pages
  \myhref{http://dx.doi.org/10.1109/SFCS.1998.743486}{362--367}, 1998.
\newblock  \myhref{http://arxiv.org/abs/quant-ph/9805006}{{\ttfamily
  arXiv:quant-ph/9805006}}.

\bibitem{zalka:GroverOptimal}
C.~Zalka.
\newblock Grover{'}s quantum searching algorithm is optimal.
\newblock {\em Physical Review A},
  60(4):\myhref{http://dx.doi.org/10.1103/PhysRevA.60.2746}{2746}, 1999.
\newblock  \myhref{http://arxiv.org/abs/quant-ph/9711070}{{\ttfamily
  arXiv:quant-ph/9711070}}.

\bibitem{zhan:treesWithHiddenStructure}
B.~Zhan, S.~Kimmel, and A.~Hassidim.
\newblock Super-polynomial quantum speed-ups for {Boolean} evaluation trees
  with hidden structure.
\newblock In {\em Proc. of 3rd ACM ITCS}, pages
  \myhref{http://dx.doi.org/10.1145/2090236.2090258}{249--265}, 2012.
\newblock  \myhref{http://arxiv.org/abs/1101.0796}{{\ttfamily
  arXiv:1101.0796}}.

\end{thebibliography}

\appendix

\section{Basics of Representation Theory}
In this appendix, we formulate the basic results in representation theory of the symmetric group used in \rf(sec:representations).
For general representation theory of finite groups, the reader may refer to~\cite{serre:representation} and~\cite{curtis:representationTheory}.  For representation theory of the symmetric group, we mostly use~\cite{sagan:symmetricGroup}.

An {\em algebra} $A$ over a field $K$ is a vector space over $K$ that is simultaneously a ring with the  identity element.  Moreover, the algebra $A$ has to satisfy the following associativity condition: $\alpha (uv) = (\alpha u)v = u(\alpha v)$ for all $\alpha\in K$ and $u,v\in A$.

The only type of algebra we use in the paper is the {\em group algebra}.  Let $G$ be a finite group.  The group algebra $KG$ is the vector space over $K$ with the elements of $G$ forming a basis.  The ring multiplication operation for the basis elements of $KG$ is inherited from the group $G$, and then uniquely extended by linearity for the remaining elements.  That is, $\sA[\sum_{g\in G} \alpha_g g]\sA[\sum_{h\in G} \beta_h h] = \sum_{g,h\in G} \alpha_g\beta_h (gh)$.

Assume that $A$ is an algebra over $K$.  A (left) $A$-module is a vector space $M$ over $K$ such that for all $u\in A$ and $m\in M$, the product $um$ is defined that satisfies the following conditions:
\[
u(m+n) = um+un,\quad (u+v)m = um + vm,\quad (uv)m = u(vm),\quad em=m,\quad (\alpha u)m = \alpha (um),
\]
for all $\alpha\in K$, $u,v\in A$, and $m,n\in M$, and $e$ is the identity element of $A$.  A {\em submodule} of $M$ is a subspace of $M$ that is closed under multiplication by the elements of $A$.
A module $M$ is called {\em irreducible} if it does not contain any submodule except for the trivial ones: $M$ itself, and the zero-dimensional subspace $\{0\}$.

\mycommand{Hom}{\mathrm{Hom}}
Assume that $M$ and $N$ are $A$-modules.  An {\em $A$-homomorphism} from $M$ to $N$ is a linear operator $\theta\colon M\to N$ that satisfies $\theta(um) = u\theta(m)$ for all $u\in A$ and $m\in M$.  
Let $\Hom(M,N)$ denote the linear space of all $A$-homomorphisms from $M$ to $N$.
If an $A$-homomorphism $\theta$ is also a linear isomorphism, then $\theta$ is called an {\em $A$-isomorphism}, and $M$ and $N$ are called $A$-isomorphic.  

A {\em direct sum} $M\oplus N$ of $M$ and $N$ as linear spaces is an $A$-module with the operation $u(m\oplus n) = um\oplus un$ for all $u\in A$, $m\in M$ and $n\in N$.

\subsection{Representations}
We only consider $\bR G$-modules, where $G$ is a finite group, and $\bR$ is the field of real numbers.
Such modules are known as (real) {\em representations}.
To define an $\bR G$-module $M$, it suffices to define the products $gu$, where $g\in G$, and $u$ is a basis element of $M$.  The operation $u\mapsto gu$ is also known as {\em group action}.
We assume that $M$ is equipped with an inner product satisfying $\ip<u,v> = \ip<gu, gv>$ for all $g\in G$ and $u,v\in M$.  Such an inner product can be always constructed~\cite[Proof of Theorem 1.5.3]{sagan:symmetricGroup}.

\begin{lem}[{Schur's Lemma, \cite[Section 2.2]{serre:representation}, \cite[Theorem 1.6.5]{sagan:symmetricGroup}}]
Assume $\theta\colon V\to W$ is an $\bR G$-homomorphism between two irreducible $\bR G$-modules $V$ and $W$. 
Then, $\theta=0$ if $V$ and $W$ are not isomorphic.  Otherwise, $\theta$ is uniquely defined up to a scalar multiplier.
\end{lem}

Maschke's theorem \cite[Theorem 1.5.3]{sagan:symmetricGroup} implies that any $\bR G$-module is decomposable into a direct sum of pairwise orthogonal irreducible $\bR G$-modules:
\begin{equation}
\label{eqn:moduleDecompose}
M = M_1 \oplus M_2 \oplus \cdots \oplus M_m.
\end{equation}
However, this decomposition is not unique.

Let $V$ be an irreducible $\bR G$-module. 
The number of components in~\rf(eqn:moduleDecompose) isomorphic to $V$ is called the {\em multiplicity} of $V$ in $M$.
Their direct sum is the {\em canonical submodule} of $M$ associated with $V$.
Both the multiplicity and the canonical submodule do not depend on the decomposition in~\rf(eqn:moduleDecompose)~\cite[Section 2.6]{serre:representation}.

Let $N$ be a direct sum of $\ell$ copies of $V$, and let $k$ be the multiplicity of $V$ in $M$.  Then, Schur's lemma implies that, in a specifically chosen basis, any $\bR G$-homomorphism from $N$ to $M$ can be given by $A\otimes I$, where $A$ is an arbitrary $k\times\ell$-matrix, and $I$ is the $d\times d$ identity matrix, where $d$ is the dimension of $V$.
In particular, the dimension of $\Hom(N,M)$ is $k\ell$.

Assume that $M$ is an $\bR G$-module and $H$ is a subgroup of $G$.  Then, $M$ can be also considered as an $\bR H$-module.  It is called the {\em restricted} module and is denoted by $M\down_{H}$.

Let $G$ and $H$ be finite groups, $M$ be an $\bR G$-module, and $N$ be an $\bR H$-module.  Then, the tensor product of $M$ and $N$ as vector spaces, $M\otimes N$, is an $\bR(G\times H$)-module with the group action defined by $(g, h)(u\otimes v) = (gu)\otimes (hv)$ for all $(g,h)\in G\times H$, $u\in M$ and $v\in N$.  
This operation is called the {\em outer tensor product}.
The resulting module is irreducible if $M$ and $N$ are irreducible, and every irreducible $\bR(G\times H)$-module can be obtained in this way~\cite[Section 3.2]{serre:representation}.

\subsection{Representations of the Symmetric Group}
\mycommand{diagram}{{\boldsymbol\lambda}}
Throughout this section, $X$ is a finite set of $n$ elements.  Let $\bN$ denote the set of positive integers.  The symmetric group on $X$ is denoted by $\bS_X$.  It consists of all permutations on $X$.  Clearly, $\bS_X$ and $\bS_Y$ are isomorphic if $|X|=|Y|$.  

A {\em partition} of $n$ is a sequence $\lambda=(\lambda_i)_{i\in\bN}$ of non-increasing non-negative integers that sum up to $n$, denoted $\lambda\vdash n$.  In particular, $\lambda$ is eventually zero, and its description is usually truncated at the first zero.
The {\em diagram} of $\lambda$ is defined as $\diagram = \{(i,j)\in\bN^2 \mid j\le \lambda_i\}$.


A {\em tableau} of shape $\lambda$, or {\em $\lambda$-tableau}, is a bijection $t\colon \diagram\to X$.
For example, if $\lambda = (3,1)$, and $X=[4]$,
\[
t =\quad  \begin{matrix} 1 & 2 & 4 \\ 3\end{matrix}
\]
is a tableau with $t(1,1) = 1$, $t(1,2) = 2$, $t(1,3) = 4$, and $t(2,1) = 3$.
For $\pi\in \bS_X$, the notation $\pi t$ denotes the composition $\pi\circ t$, which is also a tableau of shape $\lambda$.
The $i$th row of $t$ is defined by $R_i(t) = \{t(i,j)\mid (i,j)\in\diagram\}$.
The $j$th column of $t$ is $C_j(t) = \{t(i,j)\mid (i,j)\in\diagram\}$.
For each $\lambda$-tableau $t$, we define two subgroups of $\bS_X$: $R_t = \prod_i \bS_{R_i(t)}$ and $C_t = \prod_j \bS_{C_j(t)}$.

\mycommand{basis}{v}
The {\em content} of a function $f\colon X\to\bN$ is the sequence $\sA[|f^{-1}(i)|]_{i\in\bN}$.
Assume $\lambda\vdash n$.
A {\em $\lambda$-tabloid} is a function $f\colon X\to\bN$ of content $\lambda$.
The set of all $\lambda$-tabloids forms an orthonormal basis of the corresponding {\em permutation module} $M^\lambda$.
Let us, for greater clarity, denote the basis element corresponding to $f$ by $\basis_f$.
The group action on the basis elements is given by $\pi \basis_f = \basis_{f\circ \pi^{-1}}$.

For each $\lambda$-tableau $t$, denote $\basis_t = \basis_f$, where the $\lambda$-tabloid $f$ maps $x$ to $t^{-1}(x)\elem[1]$, i.e., to the number of the row of $t$ that contains $x$.  Note that $\pi\basis_t = \basis_{\pi t}$.
\mycommand{sgn}{\mathop{\mathrm{sgn}}\nolimits}
Define the element $\kappa_t$ of the group algebra $\bR\bS_X$ by
\[
\kappa_t = \sum_{\pi\in C_t} \sgn(\pi)\pi =
\prod_{j} \sC[\sum_{\pi \in \bS_{C_j(t)}} \sgn(\pi)\pi],
\]
where $\sgn(\pi)$ denotes the sign of the permutation $\pi$.
The subspace of $M^\lambda$ spanned by $\kappa_t\basis_t$, as $t$ ranges over all $\lambda$-tableaux, is an $\bR\bS_X$-submodule.  It is known as the {\em Specht module} $S^\lambda$ corresponding to $\lambda$~\cite[Proposition 2.3.5]{sagan:symmetricGroup}.
Each irreducible $\bR\bS_X$-module is isomorphic to exactly one of the Specht modules~\cite[Theorem 2.4.6]{sagan:symmetricGroup}.

\mycommand{id}{{\mathrm{id}}}
Our next aim is to give a description of $\Hom(S^\lambda, M^\mu)$ for partitions $\lambda$ and $\mu$ of $n$.
For that, it is easier to assume that $X=\diagram$.  As $\bS_X\cong \bS_\diagram$, this is without loss of generality.
In this case, the identity function $\id\colon \diagram\to\diagram$ is a valid $\lambda$-tableau.
A {\em generalised tableau of shape $\lambda$} is a function $T\colon \diagram\to\bN$.  The tableau $T$ is called {\em semi-standard} if $T(i,j+1)\ge T(i,j)$ and $T(i+1,j)> T(i,j)$ for all $i,j$ for which these expressions are defined.
\begin{thm}[{\cite[Theorem 2.10.1]{sagan:symmetricGroup}}]
\label{thm:lambdamu}
For each generalised tableau $T$ of shape $\lambda$ and content $\mu$, there exists a unique $\bR\bS_X$-homomorphism $\theta_T\colon M^\lambda\to M^\mu$ satisfying $\theta_T(\basis_\id) = \sum_{\pi\in R_\id} \pi\basis_T$.
The set of restricted homomorphisms $\sfig{\theta_T|_{S^{\lambda}}}$, where $T$ runs through the set of all semi-standard generalised tableaux of shape $\lambda$ and content $\mu$, forms a basis of $\Hom(S^\lambda, M^\mu)$.
\end{thm}

Assume $X = Y\cup Z$ is a partition.
Let $S^\lambda$, $S^\mu$ and $S^\nu$ be Specht $\bS_X$-, $\bS_Y$- and $\bS_Z$-modules, respectively.
The Littlewood-Richardson rule~\cite[Section 4.9]{sagan:symmetricGroup} gives the multiplicity of $S^\mu\otimes S^\nu$ in $S^\lambda\down_{\bS_X\times \bS_Y}$.
The multiplicity is 0 unless $\boldsymbol\mu\subseteq \diagram$.  
Now assume that $\boldsymbol\mu\subseteq \diagram$, and consider a function $f\colon \diagram\setminus\boldsymbol\mu \to \bN$.
It is known as a {\em skew tableau}.  A semi-standard skew tableau is defined as for generalised tableaux.
The multiplicity of $S^\mu\otimes S^\nu$ in $S^\lambda\down_{\bS_X\times \bS_Y}$ is equal to the number of semi-standard tableaux $f\colon \diagram\setminus\boldsymbol\mu \to \bN$ of content $\nu$ such that the content of the restriction of $f$ onto $\{(i,j)\in\bN^2\mid j\ge a\}$ is non-increasing for any $a\in\bN$.

\subsection{Johnson Association Scheme}
In this section, we apply the general theory from the previous section to the special case used in \rf(sec:representations) and prove some results from that section.

Let $N=[n]$.
The permutation $\bS_N$-module $M(N,k)$ corresponding to a partition $\mu = (n-k,k)$ is known as the {\em Johnson association scheme}.  In this case, we identify a $\mu$-tabloid $f$ with the subset $f^{-1}(2)$.  That is, we assume that $M(N,k)$ has the set of all $k$-subsets of $[n]$ as its orthonormal basis.
The tensor product $A\otimes B$ of two disjoint subsets is understood as their union.  For example,
\[
\sA[\{1\}-\{2\}]\otimes \sA[\{3\}-\{4\}] = \{1,3\} - \{1,4\} - \{2,3\} + \{2,4\}
\]
is an element of $M(N,2)$ for $n\ge 4$.

\pfstart[Proof of \rf(lem:specht)]
We aim to apply \rf(thm:lambdamu).
Let $\lambda\vdash n$, and $X=\diagram$.
If $\lambda_3>0$, or if $\lambda_3=0$ but $\lambda_2>k$, then there is no semi-standard generalised tableaux of shape $\lambda$ and content $\mu$.  
So, we shall further assume that $\lambda_3=0$, $\lambda_2 = t\le k$.
In this case, there is unique semi-standard generalised tableau $T$ of shape $\lambda$ and content $\mu$:
\begin{equation}
\label{eqn:tableauT}
T =\quad \begin{matrix}
1 & \dots & 1 & 1 & 1 & \dots & 1 & 2  & \dots & 2\\
2 &  \dots & 2
\end{matrix}\;\;,
\end{equation}
where there are $t$ occurrences of `2' in the second row, and $k-t$ occurrences in the first row.
This proves that $M(N,k) = \bigoplus_{t=0}^k S_k(N,t)$.

It is easy to see that the dimension of $M(N,k)$ is ${n\choose k}$.  As $M(N,k)$ has only one additional irreducible submodule, $S^{(n-k,k)}$, compared to $M(N,k-1)$, the dimension of $S^{(n-k,k)}$ is ${n\choose k} - {n\choose k-1}$.

By \rf(thm:lambdamu), $\Hom(S^\lambda, M^\mu)$ is 1-dimensional.  Moreover, the only (up to a scalar factor) $\bR\bS_X$-homomorphism $\theta\colon S^\lambda\to M^\mu$ maps $\kappa_\id \basis_\id$ into $\kappa_\id \sum_{\pi\in R_\id} \pi\basis_T$.
Let us analyse the last expression in more detail.  The elements of $R_\id$ permute the elements in the rows of the tableau in~\rf(eqn:tableauT), the elements of $C_\id$ permute the elements in its columns.
Let $\pi\in R_\id$, and $U=T\circ\pi^{-1}$.  Then, $U(2,j)=2$ for all $j$.  If $U(1,j)=2$ for some $j\le t$, then $\kappa_{\id} \basis_{U}=0$, because, for any $\sigma\in C_{\id}$, $\sigma \basis_U$ cancels out with $\tau\sigma \basis_U$, where $\tau$ is the transposition exchanging $(1,j)$ and $(2,j)$.
Thus, $\kappa_\id \sum_{\pi\in R_\id} \pi\basis_T$ is proportional to a linear combination of generalised tableau $U$ of the same form as $T$, where each of the first $t$ columns of $U$ contain one `1' and one `2', and some of the next $k-t$ columns contain `2'.  Moreover, the coefficient of $U$ in this linear combination is 1 if its second row contains even number of `1's, and $-1$ otherwise.

Let us now translate this to $\bR\bS_N$.
Let $a=(a_1,\dots,a_t)$ and $b=(b_1,\dots,b_t)$ be two disjoint sequences of pairwise distinct elements of $N$.  
We choose a bijection $t\colon \diagram\to N$ such that $t(1,j) = b_j$ and $t(2,j) = a_j$ for all $j\in[t]$.
In other words, we identify the positions in the tableau with integers in $N$.  In our interpretation, a generalised tableau $U$ corresponds to the set of positions labelled by `2'.  Thus, if we apply the bijection $t$ to the homomorphism $\theta$, we get that the only $\bR\bS_N$-homomorphism from $S^\lambda$ to $M^\mu$ maps the vector
\[
\sA[\{a_1\}-\{b_1\}] \otimes \sA[\{a_2\}-\{b_2\}]\cdots\otimes \sA[\{a_t\}-\{b_t\}]
\]
(corresponding to $\kappa_\id \basis_\id$) into the vector
\[
\sA[\{a_1\}-\{b_1\}]\otimes \cdots\otimes \sA[\{a_t\}-\{b_t\}] \otimes \sB[\sum_{A\subseteq N\setminus\{a_1,\dots,a_t,b_1,\dots,b_t\}\colon |A|=k-t} A ]
\]
(corresponding to $\kappa_\id \sum_{\pi\in R_\id} \pi\basis_T$).
\pfend

\pfstart[Proof of~\rf(eqn:reductionDecompose)]
Let $N = N_0\cup N_1$ be a partition, and let $S_k(N,t)$ be the unique copy of $S^{(n-t,t)}$ in $M(N,k)$.  We aim to apply the Littlewood-Richardson rule in order to get the decomposition of $S_k(N,t)\down_{\S_{N_0}\times \S_{N_1}}$ into irreducible submodules $S^\mu\otimes S^\nu$.
As before, the multiplicity is zero if $\mu_3>0$, $\nu_3>0$, $\mu_2>k$, or $\nu_2>k$.  
Thus, let us assume $\mu_3=\nu_3=0$, and $\mu_2 = t_0$, $\nu_2=t_1$ satisfy $t_0,t_1\le k$.
Thus, if there is any skew tableau satisfying the conditions of the Littlewood-Richardson rule, it must have the form
\[
\begin{matrix}
* & \dots & * & * & \dots & * &\dots & 1 & \dots & 1\\
* & \dots & * & 1 & \dots & 2
\end{matrix}\;\;,
\]
where the $*$ stand for the elements of $\boldsymbol\mu$.  (The crucial observation here is that the right-most element of the first row must be equal to 1.)
That is, the inequality $t_1 + t_2 \le t$ must hold, and in this case the multiplicity is 1.
\pfend

\end{document}